\documentclass[12pt]{iopart}
\usepackage{graphicx}
\expandafter\let\csname equation*\endcsname\relax
\expandafter\let\csname endequation*\endcsname\relax
\usepackage{amsmath}
\usepackage{mathabx}
\usepackage{cite}
\usepackage{lscape}
\usepackage{varwidth}
\usepackage{textcomp}
\usepackage{xcolor,soul}
\usepackage{color,soul}
\usepackage{algorithm}
\usepackage{algpseudocode}
\usepackage{bm}
\usepackage{esvect}
\usepackage{accents}
\usepackage{caption}
 \usepackage{textcomp}
\usepackage[symbol]{footmisc}

\newcommand*{\dt}[1]{%
  \accentset{\mbox{\large\bfseries .}}{#1}}

\begin{document}

\title[MICROSCOPE systematics]{MICROSCOPE: systematic errors}

\author{Manuel Rodrigues$^1$, Pierre Touboul$^1$ \footnote[2]{\,Deceased in February 2021}, Gilles M\'etris$^2$, Alain Robert$^3$, Oc\'eane Dhuicque$^1$, Joel Berg\'e$^1$, Yves Andr\'e$^3$, Damien Boulanger$^1$, Ratana Chhun$^1$, Bruno Christophe$^1$, Valerio Cipolla$^3$, Pascale Danto$^3$, Bernard Foulon$^1$, Pierre-Yves Guidotti$^3$ \footnote{\,Current address: AIRBUS Defence and Space, F-31402 Toulouse, France}, Emilie Hardy$^1$, Phuong-Anh Huynh$^1$, Vincent Lebat$^1$, Fran\c{c}oise Liorzou$^1$, Benjamin Pouilloux$^3$ \footnote{\,Current address: KINEIS, F-31520 Ramonville Saint-Agne, France}, Pascal Prieur$^3$, Serge Reynaud$^{4}$, Patrizia Torresi$^3$}

\address{$^1$ DPHY, ONERA, Universit\'e Paris Saclay, F-92322 Ch\^atillon, France}
\address{$^2$ Universit\'e C\^ote d{'}Azur, Observatoire de la C\^ote d'Azur, CNRS, IRD, G\'eoazur, 250 avenue Albert Einstein, F-06560 Valbonne, France}
\address{$^3$ CNES, 18 avenue E Belin, F-31401 Toulouse, France}
\address{$^{4}$ Laboratoire Kastler Brossel, UPMC-Sorbonne Universit\'e, CNRS, ENS-PSL University, Coll\`ege de France, 75252 Paris, France}

\ead{manuel.rodrigues@onera.fr, gilles.metris@oca.eu}\vspace{10pt}
\begin{indented}
\item[]April 2021
\end{indented}

\begin{abstract}
The MICROSCOPE mission aims to test the Weak Equivalence Principle (WEP) in orbit with an unprecedented precision of 10$^{-15}$ on the E\"otv\"os parameter thanks to electrostatic accelerometers on board a drag-free micro-satellite. The precision of the test is determined by statistical errors, due to the environment and instrument noises, and by systematic errors to which this paper is devoted. Systematic error sources can be divided into three categories: external perturbations, such as the residual atmospheric drag or the gravity gradient at the satellite altitude, perturbations linked to the satellite design, such as thermal or magnetic perturbations, and perturbations from the instrument internal sources. Each systematic error is evaluated or bounded in order to set a reliable upper bound on the WEP parameter estimation uncertainty. 
\end{abstract}

%
\vspace{2pc}

%
\submitto{\CQG}
%
%
%

\section{Introduction}

The MICROSCOPE mission aims to test the Weak Equivalence Principle (WEP) with an accuracy of 10$^{-15}$. In order to reach this objective, two twin electrostatic accelerometers have been mounted on board a micro-satellite derived from the Myriad line \cite{robertcqg3}. Each twin accelerometer, detailed in Ref. \cite{liorzoucqg2}, is composed of two cylindrical and concentric test-masses kept motionless in electrostatic levitation by a control loop, so that both masses are submitted to the same gravitational field. After analysis of 120 orbits corresponding to only 7\% of the data, the first results were presented in Refs. \cite{touboul17, touboul19} and showed that systematic errors were dominated by upper bound thermal effect to about $9\times{}10^{-15}$. This evaluation is improved here by a better characterisation of the instrument and of the satellite performed after Refs. \cite{touboul17, touboul19}.

\begin{figure*} [h]
\includegraphics[width=0.45 \textwidth]{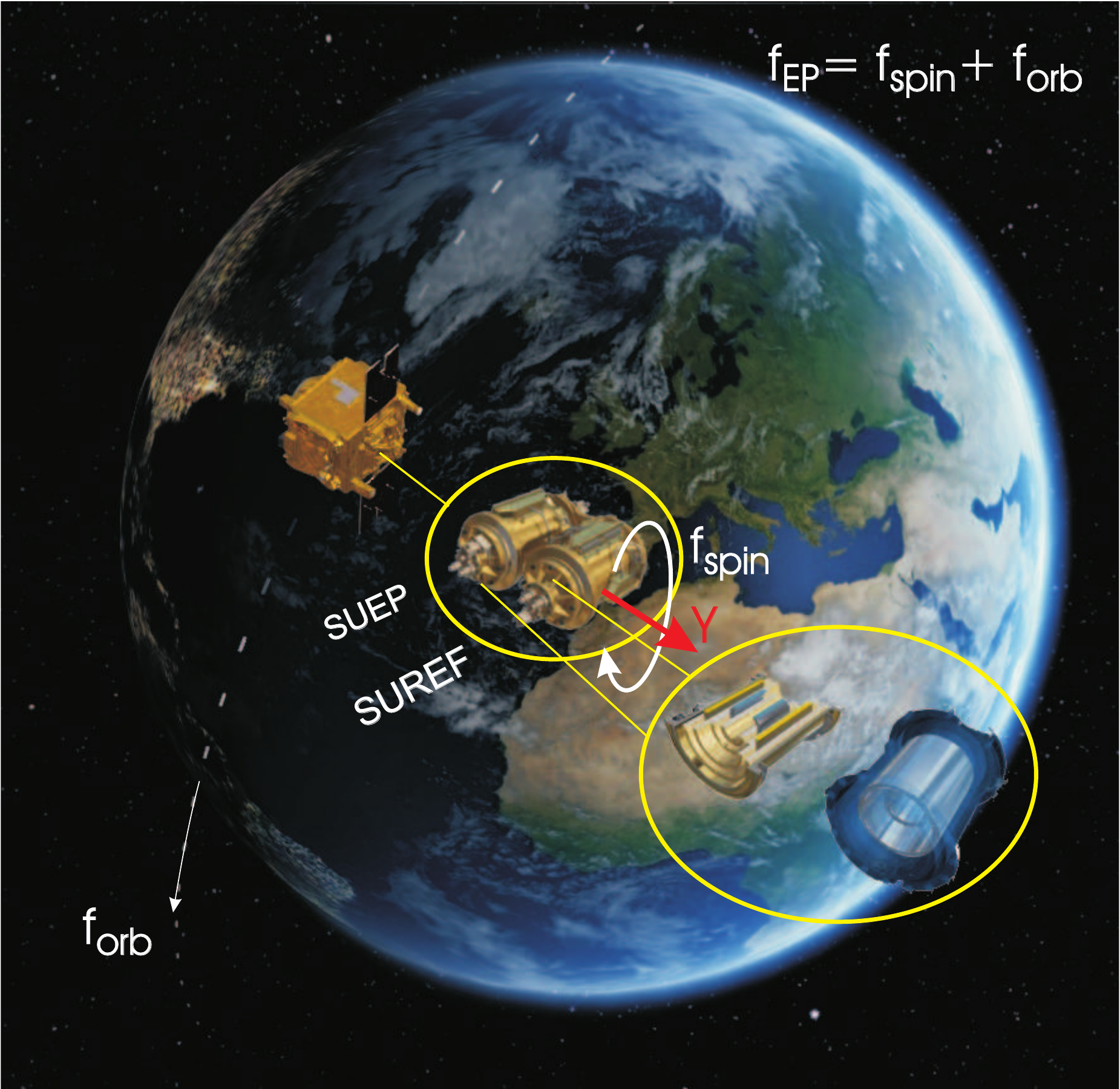}
\includegraphics[width=0.5 \textwidth]{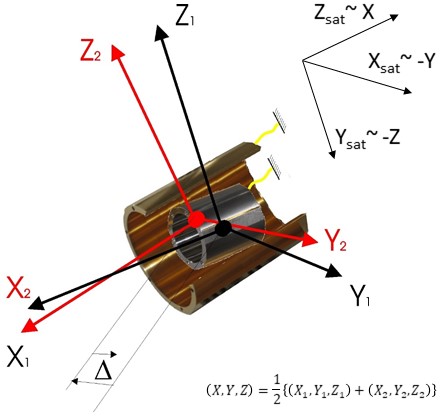}
\caption{Left: the  4 test-masses orbiting around the Earth (credits CNES / Virtual-IT 2017) . Right: test-masses and satellite frames; the ($X_{\rm sat}$, $Y_{\rm sat}$, $Z_{\rm sat}$) triad defines the satellite frame; the reference frames ($X_k$, $Y_k$, $Z_k$, $k=1,2$) are attached to the test-masses (black for the inner mass $k=1$, red for the outer mass $k=2$); the $X_k$ axes are the test-mass cylinders’ longitudinal axis and define the direction of the Equivalence Principle test measurement; the radial axis $Y_k$ and $Z_k$ are geometrically similar but connected to different sets of electrodes; during science operations, $Y_k$ axes are normal to the orbital plane, and define the rotation axis when the satellite spins. The $7\,\mu$m gold wires \cite{liorzoucqg2} connecting electrically the test-masses are shown as yellow lines. The centres of mass have been approximately identified with the origins of the corresponding sensor-cage-attached reference frames.}
\label{fig_orbit} 
\end{figure*}

A difference between the electrostatic forces applied to the test masses would indicate a violation  of the universality of free-fall and therefore of the Equivalence Principle (EP, i.e. WEP) \cite{rodriguescqg1}. The satellite \cite{robertcqg3} is placed in a sun-synchronous orbit at 710\,km altitude. It can operate either with a quasi inertial pointing or with a rotation at a frequency $f_{\rm{spin}}$ about the instrument $Y$ axis (\Fref{fig_orbit}) in the direction opposite to the orbital rotation one. The Earth's gravity field projection on $X$ axis is modulated at the frequency $f_{\rm{EP}}= f_{\rm{orb}} +f_{\rm{spin}}$, where $f_{\rm{orb}}$ is the orbital frequency: \Tref{tab_freq} gives all the frequencies of interest for different satellite modes (rotating, calibration). As shown on \Fref{fig_orbit}, the
instrument reference frame ($X_{\rm inst}$, $Y_{\rm inst}$, $Z_{\rm inst}$), noted ($X$,$Y$,$Z$) for simplicity, is related to the satellite reference frame up to a micro-rotation:
\begin{equation}
\begin {split}
& X \equiv X_{\rm inst} \approx -Z_{\rm sat},\\
& Y \equiv Y_{\rm inst} \approx +X_{\rm sat},\\
& Z \equiv Z_{\rm inst} \approx -Y_{\rm sat}.
\end{split}
\end{equation}

The measurement to be considered for each twin cylindrical accelerometer is the difference between the electrostatic accelerations applied on the inner test-mass and on the outer test-mass in order to keep them motionless at the centre of the accelerometer \cite{liorzoucqg2}. This measurement is processed by extracting the component at the EP frequency $f_{\rm{EP}}$ of the signal along the cylinder axis, $X$ axis considered as the measurement axis \cite{bergecqg7}. The measurement at this frequency contains the potential EP violation signal to be estimated which is disturbed by systematic and stochastic errors.

\begin{table}
\caption{\label{tab_freq} Main frequencies of interest.}
\begin{indented}
\item[]\begin{tabular}{@{}llll}
\br
Label & Frequency & Comment \\
\mr
$f_{\rm{orb}}$ & $0.16818\times{}10^{-3}$ Hz & Mean orbital frequency \\
\mr
$f_{\rm{spin}_2}$ & $\frac{9}{2}f_{\rm{orb}}$=$0.75681 \times{}10^{-3}$ Hz& Spin rate frequency 2 (V2 mode) \\
\mr
$f_{\rm{spin}_3}$ & $\frac{35}{2} f_{\rm{orb}}$=$2.94315 \times{}10^{-3}$ Hz& Spin rate frequency 3 (V3 mode) \\
\mr
$f_{\rm{EP}_2}$ & $0.92499 \times{}10^{-3}$ Hz& EP frequency in V2 mode  \\
\mr
$f_{\rm{EP}_3}$ & $3.11133 \times{}10^{-3}$ Hz& EP frequency in V3 mode \\
\mr
$f_{\rm cal}$ & $1.22848 \times{}10^{-3}$ Hz& Calibration frequency \\
\br
\end{tabular}
\end{indented}
\end{table}

The stochastic error is evaluated in Ref. \cite{metriscqg9} as the $1\sigma$ statistic error of a least-square regression of the data. It is due to the intrinsic noise of the instrument in its environment such as thermal noise, electronic noise and electronic parasitic forces integrated over the entire measurement time. In contrast, systematic errors cannot be reduced with time integration, and must be evaluated separately in order to either determine an upper bound of the error bar or correct the disturbing effect in the measurement data thanks to calibration.

In this paper, we provide a review as exhaustive as possible of the systematic errors in the MICROSCOPE measurement. These errors arise from environmental perturbations. The experiment was designed with the rejection of these external sources in mind: the Drag-Free and Attitude Control System (DFACS) of the satellite commands the cold gas propulsion system to provide the instrument with an environment as close to a free-fall motion as possible, reducing the external perturbations \cite{robertcqg3}. Moreover, since a differential measurement is used, all external perturbations impacting both test-masses should eliminated. However, tiny instrumental defects prevent them from being completely nullified, so that can still impact the measurement. The mission accuracy can be improved by taking into account their impact on the scientific measurement with the evaluation of the different instrumental parameters, through calibration, as they are detailed in the measurement equation in \sref{inst_sys}. Some of these parameters can be evaluated in orbit, enabling us to correct part of the perturbations impact on the measurement. Other perturbations are linked to the satellite design and their levels are verified with respect to the specifications \cite{rodriguescqg1}. In \sref{therm_sys}, the impact of the thermal perturbations is evaluated. Because the signal has been measured over different sessions over two years, it is necessary to take into account the thermal stability of the experiment. 
\Sref{magn_sys} is dedicated to the analysis of the magnetic sensitivity of the experiment and the impact of the electromagnetic perturbations on the accuracy of the EP parameter estimation. Finally, the local gravity effects, corresponding to the impact of the gravity field generated locally by the satellite, is estimated in \sref{grav_sys}. \Sref{sum_sys} summarises all the systematics and provides the global budget for systematic errors affecting the estimation of the EP parameter’s estimation.

\section{Instrument systematics} \label{inst_sys}

The MICROSCOPE mission carries two differential electrostatic accelerometers developed by ONERA (see Ref. \cite{liorzoucqg2} for a detailed description). Each instrument is composed of two cylindrical and concentric test-masses, made of different material for the sensor unit dedicated to the WEP test (SUEP), and of the same material for the second sensor unit (SUREF) which intends to verify the good behaviour of the experiment. The two test-masses of each sensor unit are kept in electrostatic levitation by a set of six pairs of electrodes composing a capacitive detector, thanks to which the position of each test-mass is measured along its six degrees of freedom (three rotations and three translations). Each sensor unit is driven by a front end electronic unit (FEEU) that generates all the low noise voltages to be applied on the control electrodes. The position measurement is picked up in the FEEU and converted into a number to be transmitted to the digital servo-control. The same electrodes are used to apply the voltage computed by a control loop to compensate for the test-mass motion and keep it at the centre of its electrostatic cage. The two test-masses follow the same orbit and are submitted to the same gravitational field. A difference between the applied electrostatic accelerations would therefore indicate a violation of the WEP.

However, the measurement of this differential acceleration is impacted by errors due to geometrical imperfections of the instrument and perturbative accelerations. Some of these perturbations depend on parameters that can be calibrated in-flight in order to minimise their effect on the measurement.

\subsection{Impact of the instrument defects} \label{sect_def}
The ideal measurement of the accelerometer is the electrostatic acceleration applied by the electrodes on the test-mass in order to keep it motionless with respect to the satellite. However, the measurement is not direct: the electrostatic acceleration is inferred from the voltages applied on the electrodes. In practice, the measurement is impacted by an offset, test-masses offcentrings, scale factor knowledge accuracy, couplings between axes, rotations of the test-masses frames with respect to the mean instrumental reference frame in which the measurement is expressed, couplings with the angular acceleration, quadratic terms and noises.
 
The difference of acceleration, called differential acceleration for simplification in this paper, to be evaluated for the test is described by the measurement equation detailed in Ref. \cite{rodriguescqg1} and summarised in \Eref{eq_xacc}. For a given quantity $s$, the notation $s^{(d)}$ refers to the difference between the inner and the outer test-mass signals: $s$ being an acceleration, a bias or a noise. The notation $\beta_{d}$ refers to the half difference between the inner and the outer test-mass parameters: $\beta$ being a sensitivity matrix or one of its component,  a quadratic parameter or a coupling factor. Similarly $s^{(c)}$ and $\beta_{c}$ refer to the common-mode signals and parameters, that is to say the half sum of their values for the inner test-mass and the outer test-mass. The differential acceleration measurement is considered along the $X$ axis reads as:

\begin{equation}  \label{eq_xacc}
\begin{split}
\Gamma_x^{(d)} &\approx  B_{0x}^{(d)} \\
    &+  \tilde{a}_{c11} b_{1x}^{(d)} + \tilde{a}_{c12} b_{1y}^{(d)}+ \tilde{a}_{c13} b_{1z}^{(d)} \\
    &+ \tilde{a}_{c11} \, \delta \, g_x + \tilde{a}_{c12} \, \delta \, g_y + \tilde{a}_{c13}\, \delta \, g_z \\
    & + \left(T_{xx} - {\rm In}_{xx} \right) \tilde{a}_{c11} \Delta_x + \left(T_{xy} - {\rm In}_{xy} \right) \tilde{a}_{c11} \Delta_y + \left(T_{xz} - {\rm In}_{xz} \right) \tilde{a}_{c11} \Delta_z \\
    & + \left(T_{yx} - {\rm In}_{yx} \right) \tilde{a}_{c12} \Delta_x + \left(T_{yy} - {\rm In}_{yy} \right) \tilde{a}_{c12} \Delta_y + \left(T_{yz} - {\rm In}_{yz} \right) \tilde{a}_{c12} \Delta_z \\
    & + \left(T_{zx} - {\rm In}_{zx} \right) \tilde{a}_{c13} \Delta_x + \left(T_{zy} - {\rm In}_{zy} \right) \tilde{a}_{c13} \Delta_y  + \left(T_{zz} - {\rm In}_{zz} \right) \tilde{a}_{c13} \Delta_z \\
    &+ 2\left( \frac{a_{d11}}{a_{c11}} {\Gamma}_x^{(c)} + \frac{a_{d12}}{a_{c22}}{\Gamma}_y^{(c)} + \frac{a_{d13}}{a_{c33}} {\Gamma}_z^{(c)}  \right)\\ 
    &- 2\left[\tilde{a}_{c11} \left(-\Omega_z \dot{\Delta}_y+ \Omega_y \dot{\Delta}_z\right)+\tilde{a}_{c12} \left(\Omega_z \dot{\Delta}_x- \Omega_x \dot{\Delta}_z\right)+\tilde{a}_{c13} \left(-\Omega_y \dot{\Delta}_x+ \Omega_x \dot{\Delta}_y\right)\right]\\
    &-\tilde{a}_{c11}\ddot{\Delta}_x-\tilde{a}_{c12}\ddot{\Delta}_y-\tilde{a}_{c13}\ddot{\Delta}_z\\
    &+ 2 \left(c'_{d11} \dt{\Omega}_x + c'_{d12} \dt{\Omega}_y + c'_{d13} \dt{\Omega}_z  \right)\\
    &+ n_x^{(d)}-2\left( \frac{a_{d11}}{a_{c11}} n_x^{(c)} + \frac{a_{d12}}{a_{c22}} n_y^{(c)} + \frac{a_{d13}}{a_{c33}} n_z^{(c)}  \right) \\
    &+\Gamma_{q,x}^{(d)} ,\\
\end{split}
\end{equation}

with:
\begin{itemize}
\item $\left[\tilde{a}_{c11}, \tilde{a}_{c12}, \tilde{a}_{c13}\right]$ the first row of the (approximated) common-mode sensitivity matrix that takes into account the scale factors, the linear couplings between the axes and the test-masses’ reference frames misalignment. In particular, the common-mode sensitivity matrix introduces in the measurement the effect of the bias projection in the instrumental reference frame. $\tilde{a}_{c11}$ is an effective parameter entering in the measurement equation, close to  ${a}_{c11}$ but including also second order terms depending on others, ${a}_{cij}$ and ${a}_{dij}$. Similarly the other $\tilde{\,}$ symbols refer to effective parameters.

\item $\vv{B_0}^{(d)}$ the differential ``electrical bias'' of the instrument. The electrical bias is mainly a continuous (DC) signal which can vary at $f_{\rm{EP}}$ with the temperature sensitivity and can slowly drift over time. The random variations are included in the $n^{(d)}$ term.

\item $\vv{b_1}^{(d)}$ the differential ``mechanical bias'' corresponding to acceleration perturbations applied on the test-masses: the gravitational force applied by the satellite and the non-gravitational accelerations, including the electrostatic perturbations. These accelerations can also drift over time and vary at $f_{\rm{EP}}$. Part of the contributors will be evaluated in the following sections. In $\vv{b_1}^{(d)}$, the component linked to the electrostatic parasitic accelerations is mostly due to the stiffness of the sensors: the motion of a test-mass relatively to its electrodes cage induces an acceleration proportional to the displacement, resulting in a measurement offset. The stiffness values were evaluated during the commissioning phase \cite{chhuncqg5} and found to be higher than expected. However, the test-masses are servo-controlled during the WEP test sessions, and the effect of their motion at $f_{\rm{EP}}$ can therefore be limited to values lower than $10^{-15}$\,ms$^{-2}$. The stiffness also contributes to the stochastic error but remains a small contributor to the instrumental noises. 

\item $\delta$ the E\"otv\"os parameter, to be determined and  $\vv{g}$ the Earth gravity acceleration. $\delta \vv{g}$ is projected on the measurement axis through the common-mode sensitivity matrix.

\item $\vv{\Delta}$ the so-called ``offcentring'' corresponding to the vector between the two test-masses’ centres. It introduces in the measurement the effect of the Earth gravity gradient ($[T]$ the corresponding matrix and $T_{ij}$ its components in the instrument reference frame) and of the inertia gradient matrix corresponding to the rotation of the satellite in the instrument reference frame ($[{\rm In}]$ corresponding matrix and ${\rm In}_{ij}$ its components) \cite{rodriguescqg1, touboul19}.   

\item $\dot{\vv{\Delta}}$ and $\ddot{\vv{\Delta}}$ the first and second time derivates of the offcentring, and $\vv{\Omega}$ the angular velocity of the satellite, which contribute to the terms corresponding to the projection of the Coriolis effect upon the measurement axes and to the motion of the test-mass relatively to the satellite. During the EP test sessions, the electrostatic control loop keeps both test-masses motionless, and those two terms are therefore negligible. However, they have to be considered during the calibration session aimed to estimate the parameters of the common-mode sensitivity matrix. 

\item $\vv{\Gamma}^{(c)}$ the common-mode measurement acceleration, which can be minimised thanks to DFACS system compensating external accelerations. Its residual is mostly eliminated in the differential mode, but nonetheless impacts the measurement through the first line of the differential sensitivity matrix  $\left[{a}_{d11}, {a}_{d12}, {a}_{d13}\right]$.

\item $\left[{c'}_{d11}, {c'}_{d12}, {c'}_{d13}\right]$ the first row of the differential angular to linear coupling matrix, projecting the angular acceleration of the satellite $\dot{\vv{\Omega}}$ on the linear measurement.

\item $n_i^{(d)}$ and $n_i^{(c)}$ the differential and mean instrumental noises along $i$ axis.

\item $\Gamma_{q,x}^{(d)}$ is the non-linear part described bellow.
\end{itemize}

 The projection along $X$ of the differential acceleration measurement described in Ref. \cite{rodriguescqg1} neglected non-linearities. One way to consider non-linearities is to add a quadratic term $\vv\Gamma_{q}^{(d)}$ to the measurement equation:
\begin {equation}
\vv\Gamma_{q}^{(d)} = \left[ {\mathbf K_{21}}\right] \left(\vv{\gamma}^{(1)} \circ \vv{\gamma}^{(1)} \right)- \left[ {\mathbf K_{22}}\right] \left(\vv{\gamma}^{(2)} \circ \vv{\gamma}^{(2)} \right),
\end {equation}
where $\left[ {\mathbf K_{21}}\right]$ is the matrix of quadratic parameters of test-mass 1, $\left[ {\mathbf K_{22}}\right]$ being the one of test-mass 2. The applied accelerations are $\vv{\gamma}^{(i)}$ on the $i-$th test-mass. The symbol $\circ$ is the Hadamard product operator. By considering the applied differential acceleration $\vv{\gamma}^{(d)}$, the mean applied acceleration $\vv{\gamma}^{(c)}$, and the definition of the differential and common mode quadratic parameter matrices, $\left[ {\mathbf K_{2d}}\right]=1/2\left(\left[ {\mathbf K_{21}}\right]- \left[ {\mathbf K_{22}}\right]\right)$ and $\left[ {\mathbf K_{2c}}\right]=1/2\left(\left[ {\mathbf K_{21}}\right]+ \left[ {\mathbf K_{22}}\right]\right)$, the differential quadratic term included now the measurement equation \Eref{eq_xacc} is: 
\begin {equation} \label{eq_q}
\Gamma_{q,x}^{(d)} =2\left[K_{2c,xx}\gamma_x^{(c)}\gamma_x^{(d)}+K_{2d,xx}\left({\gamma_x^{(c)}}^2+\frac{1}{4} {\gamma_x^{(d)}}^2\right)\right]+O(\gamma_y^{(d)},\gamma_z^{(d)}) ;
\end {equation}
the remaining terms $O(\gamma_y^{(d)},\gamma_z^{(d)})$ containing $K_{2c,xy}$ and $K_{2c,xz}$ terms have been linearised and modelled by the coupling term $c'_d$. The quadratic factors $K_{21,xx}$ and $K_{22,xx}$ for each test-mass were estimated for each sensor during the instrument characterisation sessions in Ref. \cite{chhuncqg5}. As shown in \sref{calib}, the term $K_{2d,xx}=\frac{1}{2}(K_{21,xx}-K_{22,xx})$ can also be estimated directly from the calibration sessions of $a_{d11}$ as a byproduct. In the a priori error budget, the mean common applied acceleration $2.5\times{}10^{-8}$\,m\,s$^{-2}$ has been considered.

In \Eref{eq_xacc}, we look for the projection a signal $\delta \, \vv{g}$ collinear to the Earth's gravity field vector. To achieve the accuracy objective of the mission, it is then necessary to measure a differential acceleration as low as $7.9\times{}10^{-15}$\,m\,s$^{-2}$ at $f_{\rm EP}$, the frequency modulation of the Earth's gravity field whose amplitude is 7.9\,m\,s$^{-2}$ at the satellite altitude of 710\,km. 

All other terms in \Eref{eq_xacc} are perturbations that reduce the accuracy of the estimation. Some of them have limited effects because they have been specified to a sufficient level \cite{rodriguescqg1}. Others can be calibrated in flight in order to correct the measurements. The offcentring calibration described in the following sections has been simplified. In fact, the calibration process estimates a quantity $\tilde{a}_{c11}\Delta_{xs}$ where $\Delta_{xs}$ is the offcentring along $X$ in the satellite reference frame. In the same way, $\tilde{a}_{c11}\Delta_{zs}+\tilde{a}_{c13}\Delta_{xs}$ is calibrated for the $Z$ axis and  $\tilde{a}_{c11}\Delta_{ys}+\tilde{a}_{c12}\Delta_{xs}$ for the $Y$ axis. As those terms are also the ones that disturb the differential acceleration measurement, when the correction is applied, it is not necessary to know exactly the terms $\tilde{a}_{cij}$. They are already included in the estimator. For simplicity, we use the terms $\tilde{a}_{cij}\Delta_{j}$ for the offcentring implying that $\tilde{a}_{cij}$ do not need to be estimated.

The maximal contribution of the perturbations before calibration are summarised in \Tref{tab_param} in inertial pointing and in spin mode. It must be noted that only spin mode was used for the EP test, so that the column of the inertial pointing is only informative. The calibration aims to estimate the offcentring to an objective accuracy of $0.2\mu$m along $X$ and $Z$ and of $2\mu$m along $Y$ and the $a_{d1m}$ ($m={1,2,3}$) parameters to an objective accuracy of $10^{-4}$. After correction of the calibrated parameters, all the terms of \Tref{tab_param} are reduced by a factor 100 or 10 (except the quadratic term not used for correction of the measurement) leading to a total of $3\times{}10^{-15}$\,m\,s$^{-2}$ expected systematic effect. Each session has been corrected with calibrated parameters and the results are presented at the end of this paper in good agreement with this first error estimation.

\begin{table}
\caption{\label{tab_param} Upper bound a priori systematic effect for a set of error sources {\bf before calibration}. In this budget, the residual acceleration $ {{\Gamma}_{\rm resdf}}$ provided by the DFACS is specified \cite{rodriguescqg1} lower than $10^{-12}$\,m\,s$^{-2}$ at $f _{\rm{EP}}$.} 
\begin{indented}
\item[]\begin{tabular}{@{}lcccc}
\br
Perturbative term & Limiting parameter  &  \multicolumn{2}{c}{Acceleration}  \cr
                         & on-ground estimation & \multicolumn{2}{c}{impact at $f _{\rm{EP}}$ in m\,s$^{-2}$} \cr
                         & & Inertial  & Spin V2 / V3 \cr
\br
$T_{xx} \tilde{a}_{c11} \Delta_x $ & $\tilde{a}_{c11} \Delta_x<20.2\mu$m & $84\times{}10^{-15}$ & $0.3\times{}10^{-15}$\cr
\mr
$T_{xz} \tilde{a}_{c11} \Delta_z $ & $\tilde{a}_{c11} \Delta_z<20.2\mu$m & $86\times{}10^{-15}$ &$0.3\times{}10^{-15}$ \cr
\mr
$T_{xy} \tilde{a}_{c11} \Delta_y $ & $\tilde{a}_{c11} \Delta_y<20.2\mu$m & $0.6\times{}10^{-15}$ & $0.1\times{}10^{-15}$\cr
\mr
$T_{yy} \tilde{a}_{c12} \Delta_y $ & $\tilde{a}_{c12} < 2.6 \times{}10^{-3}$\,rad & $0.9\times{}10^{-15}$ &$0.0\times{}10^{-15}$\cr
                                                 & $\Delta_y<20\mu$m & \cr
\mr
$T_{zz} \tilde{a}_{c13} \Delta_z $ & $\tilde{a}_{c13} < 2.6 \times{}10^{-3}$\,rad & $0.7\times{}10^{-15}$ &$0.0\times{}10^{-15}$\cr
                                                 & $\Delta_z<20\mu$m & &\cr
\mr
$2\frac{a_{d11}}{a_{c11}} {{{\Gamma}_{\rm resdf}}_{,x}} $ & $\frac{a_{d11}}{a_{c11}} <10^{-2}$ &  \multicolumn{2}{c}{$20\times{}10^{-15}$}  \cr
\mr
$2\frac{a_{d12}}{a_{c11}} {{{\Gamma}_{\rm resdf}}_{,y}} $ & $\frac{a_{d12}}{a_{c11}} <1.6\times{}10^{-3}$\,rad &  \multicolumn{2}{c}{$3\times{}10^{-15}$} \cr
\mr
$2\frac{a_{d13}}{a_{c11}} {{{\Gamma}_{\rm resdf}}_{,z}} $ & $\frac{a_{d13}}{a_{c11}} <1.6\times{}10^{-3}$\,rad  &  \multicolumn{2}{c}{$3\times{}10^{-15}$} \cr
\mr
$4K_{2c,xx}(\gamma_{c,x} {{{\Gamma}_{\rm resdf}}_{,x}}) $ & $K_{2,cxx}<14000$\,s$^2$\,m$^{-1}$  &  \multicolumn{2}{c}{$1.4\times{}10^{-15}$} \cr
\mr
$2K_{2d,xx}(\gamma_{c,x}^2 +{{{\Gamma}_{\rm resdf}}_{,x}}^2) $ & $K_{2,dxx}<12000$\,s$^2$\,m$^{-1}$  &  \multicolumn{2}{c}{$1.2\times{}10^{-15}$} \cr
\br
Total in m\,s$^{-2}$ & & $200.8\times{}10^{-15}$ & $29.2\times{}10^{-15}$ \cr
\multicolumn{2}{l}{in E\"otv\"os parameter accuracy} & $25.4\times{}10^{-15}$ & $3.7\times{}10^{-15}$\cr
\br
\end{tabular}
\end{indented}
\end{table}

\subsection{In-orbit estimation of the instrument defects} \label{Sect_Off}

The in-flight calibration consists of applying to the satellite (through the drag-free loop) or to the test-masses (through the accelerometer control loop) imposed motions that amplify the effect of the parameter to be estimated, so that the corresponding effect dominates the measurement equation at a given frequency ($f_{\rm cal}$ in \Tref{tab_freq}) decorrelated from any natural frequency. The calibration protocol was validated through numerical simulation in order to verify that the estimation accuracy of the instrumental parameters can be achieved \cite{hardy13a}. 
The calibration sessions have been performed regularly in order to observe any potential time drift of the calibrated parameters used to correct the EP session data spread over the two-year mission \cite{rodriguescqg4}. 

\subsubsection{Offcentring along the $X$ and $Z$ axes}
Because the positions of the centres of mass of the inner test-mass and of the outer test-mass are not exactly the same, the gravity field as well as the inertia and Coriolis effects are slightly different between the two test-masses, thus impacting the differential measurement. As shown in \Eref{eq_xacc}, the offcentring $\vv{\Delta}$ between the two test-masses introduces effects of the Earth gravity gradient matrix $[T]$ and of the gradient of inertia matrix ${\rm [In]}$. The gravity gradient signal is modulated in the instrument reference frame by the satellite rotation around the Earth and therefore presents an important amplitude at 2$f _{\rm{EP}}$ along the $X$ and $Z$ axes. This signal is accurately computed thanks to the ITG-GRACE2s model \cite{mayer06} coming from the GRACE space mission and thus particularly adequate for the calibration.

\begin{figure*}
\includegraphics[width=0.5 \textwidth]{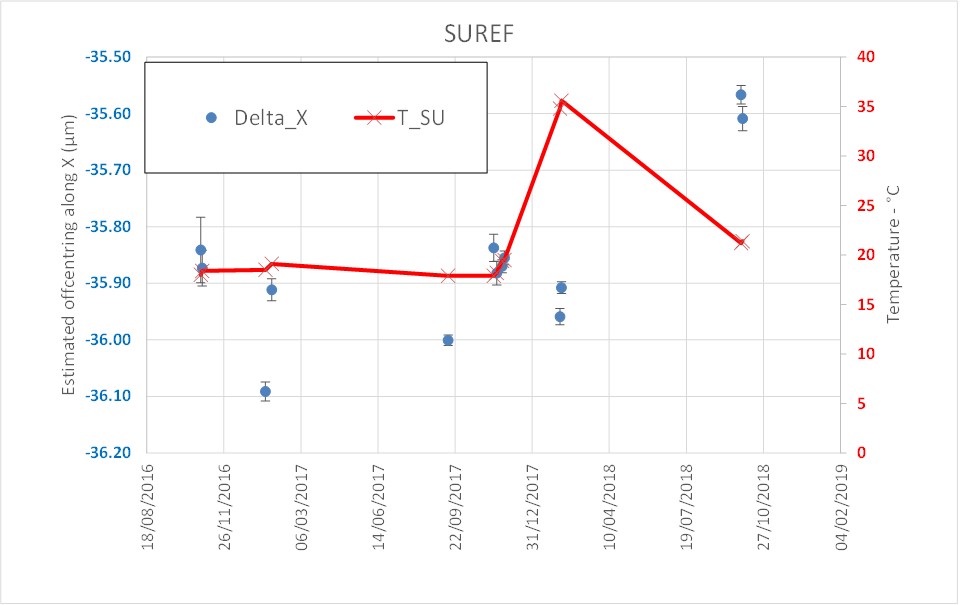}
\includegraphics[width=0.5 \textwidth]{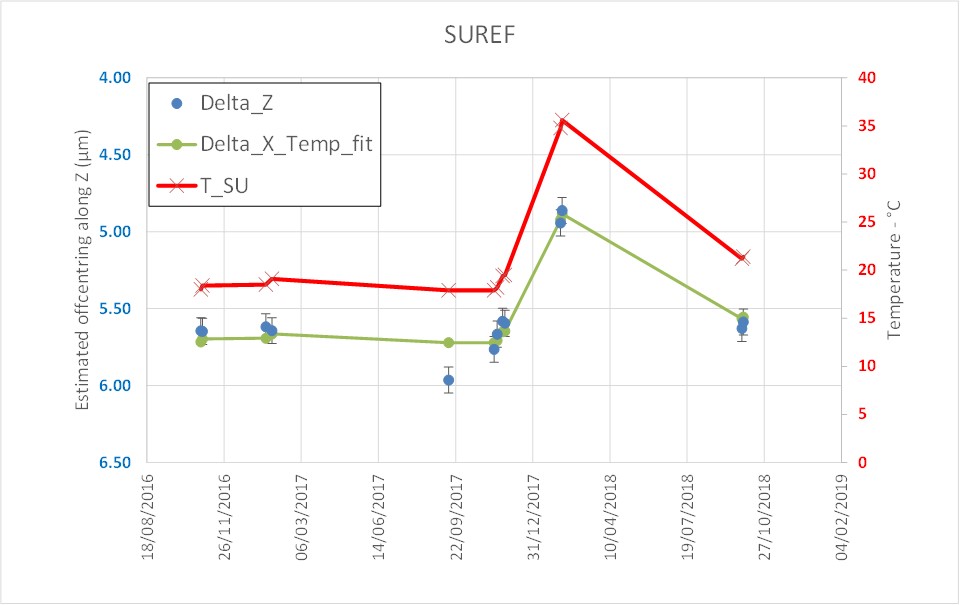}
\includegraphics[width=0.5 \textwidth]{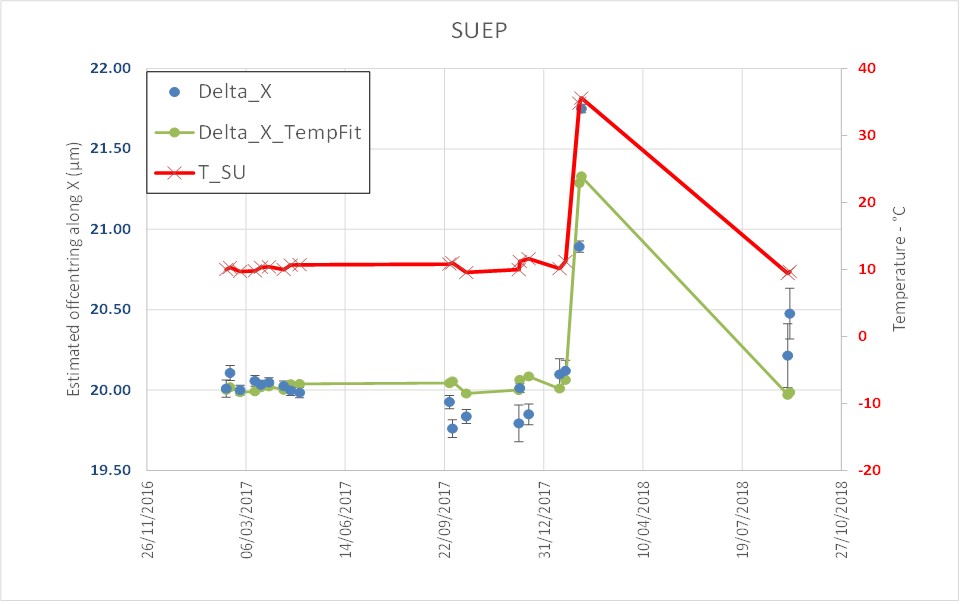}
\includegraphics[width=0.5 \textwidth]{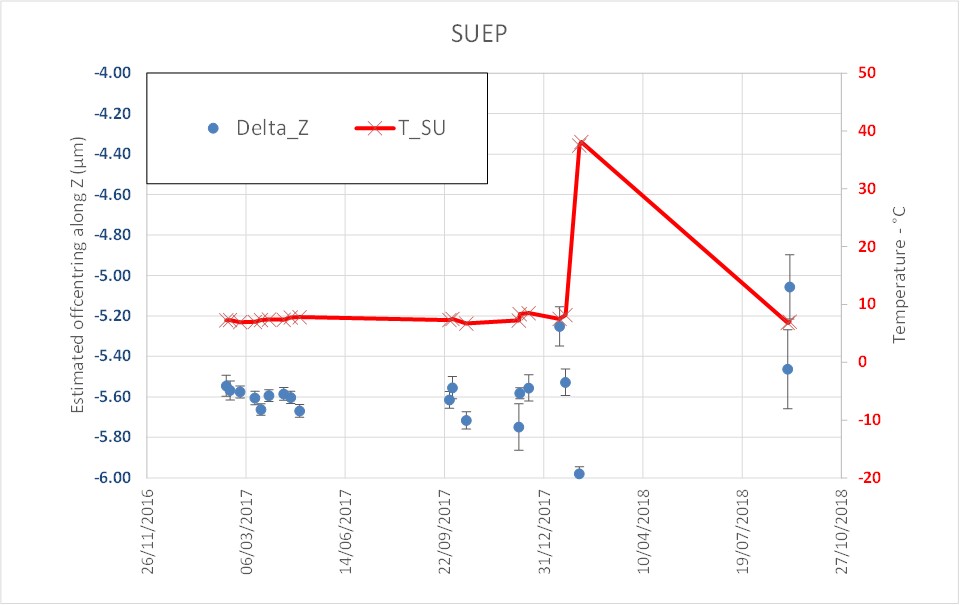}
\caption{SUREF (upper panels) and SUEP (lower panels) offcentring calibration along $X$ axis (left) and along $Z$ (right) and temperature monitoring. The green lines (Temp fit) represent the linear model fitting the parameter to the temperature when significant.}
\label{fig_Off} 
\end{figure*}

The offcentring calibration along $X$ and $Z$ does not require any specific motion of the satellite or of the test-masses; they are estimated using the measurement of the EP test sessions at the same time as the E\"otv\"os parameter. 
The resulting estimations are plotted in \Fref{fig_Off} for both SU with the mean temperature evaluated during the session. It must be noted that offcentrings are estimated during scientific sessions and thus only on limited periods of the year, typically from autumn to the end of spring. A small correlation with temperature is observed that allowed to estimate a sensitivity lower than 0.05$\mu$m\,\textdegree{}C$^{-1}$ for SUEP and lower than 0.04$\mu$m\,\textdegree{}C$^{-1}$ for SUREF when significant. It is mostly dominated by the effect of the thermal expansion of the silica parts that carry the electrodes. An estimation of the maximum sensitivity can be computed to $\pm{}0.02\mu$m\,\textdegree{}C$^{-1}$ for the inner test-mass and to $\pm{}0.03\mu$m\,\textdegree{}C$^{-1}$ with the temperature considered at the electrode location. But the temperature is measured at the mechanical interface of the SU and can be filtered. The effect of the test-mass expansion has to be added but with smaller amplitudes since the temperature is more filtered as the test-mass is only thermally coupled by radiative effects. Finally, only thermal differential effect on test-mass centres have to be considered. The observed electrode dissymmetries in Ref. \cite{chhuncqg5} lead to capacitance biases and thus relative offcentrings of the two test-masses when they are controlled by the electrodes. In the case of SUREF, the main offcentring budget comes from the inner test-mass control and for SUEP from the outer test-mass control. The inner electrode temperature is more filtered and thus leads to an apparent lower offcentring thermal sensitivity for SUREF. It is not the unique source of sensitivity: the plate holding the two sensor units undergoes mechanical deformations that induce small distortions of the parts and thus small displacements observed at this level of sensitivity. In conclusion, the a priori budget on each test-mass allows to evaluate an approximate sensitivity value of $\pm{}0.05\mu$m\,\textdegree{}C$^{-1}$ but it is rather difficult to establish for each axis and each test-mass a better guess. A correction of the offcentring with temperature could be envisaged, but is not necessary as they are estimated for each session at the same time as the E\"otv\"os parameter with sufficient accuracy. 

\begin{table}
\caption{{\label{tab_off}}Summary of offcentring calibration in a stationary temperature range}
\begin{tabular}{lccc} 
\textbf{Sensor} & \multicolumn{2}{c}{\textbf{Offcentring ($\pm{}$ standard deviation)}} & \textbf{Temperature domain}        \\ 
& Along X&Along Z&\\ \hline
\textbf{SUREF} & -35.884\,$(\pm{}0.005)\, \mu$m &  5.712\,$(\pm{}0.005)\, \mu$m  & [+17.9\textdegree{}C ; +21.4\textdegree{}C] \\
\textbf{SUEP}   & 19.998\,$(\pm{}0.009)\, \mu$m &  -5.605\,$(\pm{}0.009)\, \mu$m  & [+9.5\textdegree{}C ; +11.6\textdegree{}C] \\
\end{tabular}
\end{table}

The mean values of the estimated parameters are presented in \Tref{tab_off} within a limited range of temperature. The few sessions performed at higher temperature, in early 2018, are nevertheless taken into account in the final budget presented in \Sref{sum_sys}. The calibration accuracy of the offcentring  is better than 0.01\,$\mu$m whatever the SU in the nominal temperature range and shows a good stability over the two years of mission. Most of the estimations shown in \Fref{fig_Off} are performed with an accuracy one order of magnitude better than the required value of $0.2\mu$m as considered in \Tref{tab_param} and in Ref. \cite{hardy13a}.

\subsubsection{Offcentring along the $Y$ axis}
\begin{figure*}
\includegraphics[width=0.5 \textwidth]{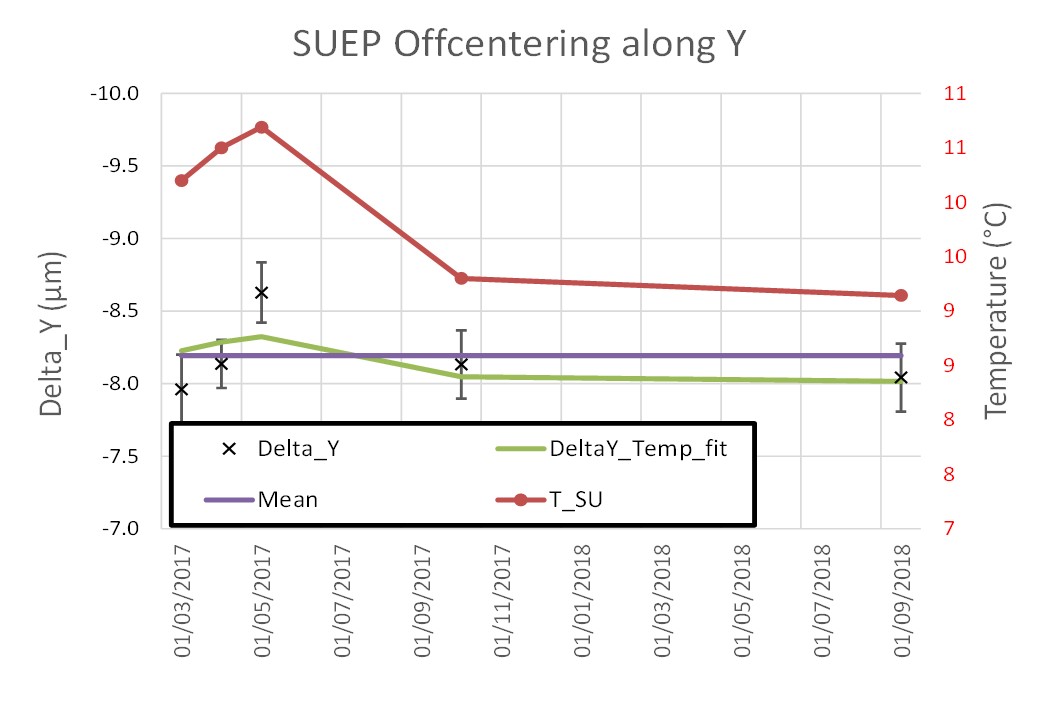}
\includegraphics[width=0.5 \textwidth]{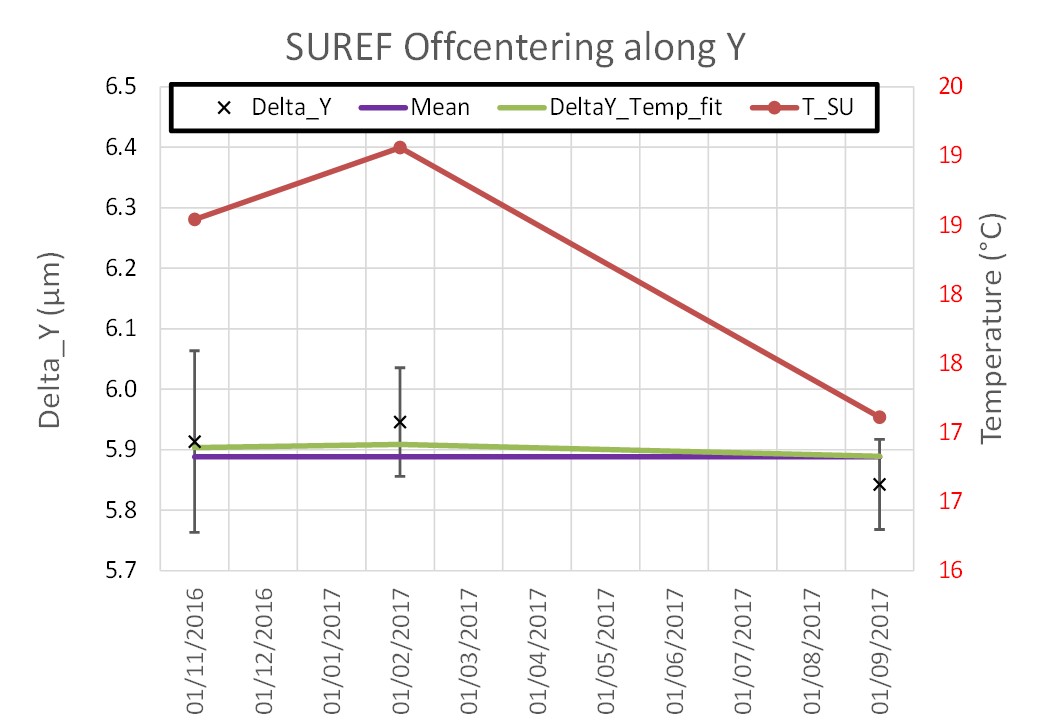}
\caption{In-orbit estimation of the offcentring along the $Y$ axis for the SUREF (right) and for the SUEP (left) with temperature monitoring. The green lines (Temp fit) represent the linear model fitting the parameter to the temperature when significant.}
\label{fig_Off_y} 
\end{figure*}

The estimation method used to estimate the offcentring along the $X$ and $Z$ axes cannot be applied for the estimation of $\Delta_y$. The $Y$ axis is orthogonal to the orbital plane, and therefore the corresponding component of the gravity gradient $T_{xy}$ is too weak to discriminate the signal from the measurement. In order to amplify its effect, the satellite is oscillated around the $Z$ axis with an amplitude of 0.05\,rad at $f_{\rm cal}$, thanks to the DFACS.
The resulting estimations are summarised in \Fref{fig_Off_y}. The mean values (and standard deviation) of the offcentring estimations are $\Delta_y=5.89(\pm{}0.05)\,\mu$m for SUREF and $\Delta_y=-8.19(\pm{}0.09)\,\mu$m for SUEP. The sensitivity to temperature is lower than 0.2$\mu$m\,K$^{-1}$ for SUEP. The sensitivity is not produced for SUREF because of too few estimations making it not relevant.

		\subsubsection{Differential scale factors, alignments and non-linearity} \label{calib}
The components of the first row of the differential sensitivity matrix $\left[a_{d11}, a_{d12}, a_{d13} \right]$ quantify the projection of the effect of the common mode acceleration into the differential measurement. $a_{d11}$ is the differential scale factor while $a_{d12}$ and $a_{d13}$ include the difference of misalignments and the couplings between the instrument axes. 

\begin{figure*}
\includegraphics[width=0.5 \textwidth]{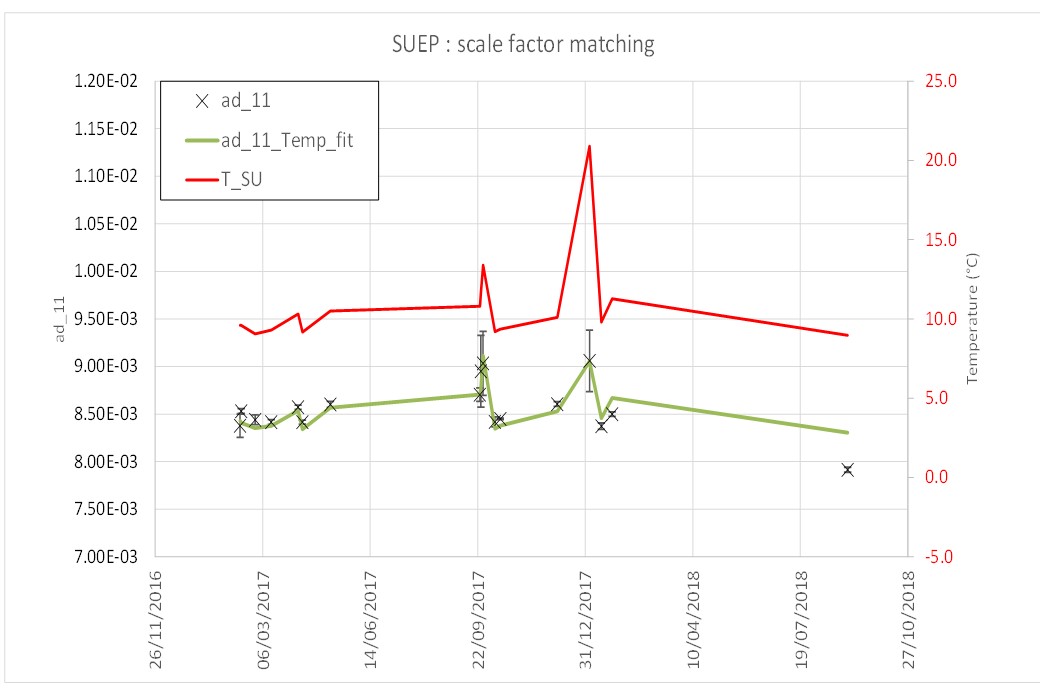}
\includegraphics[width=0.5 \textwidth]{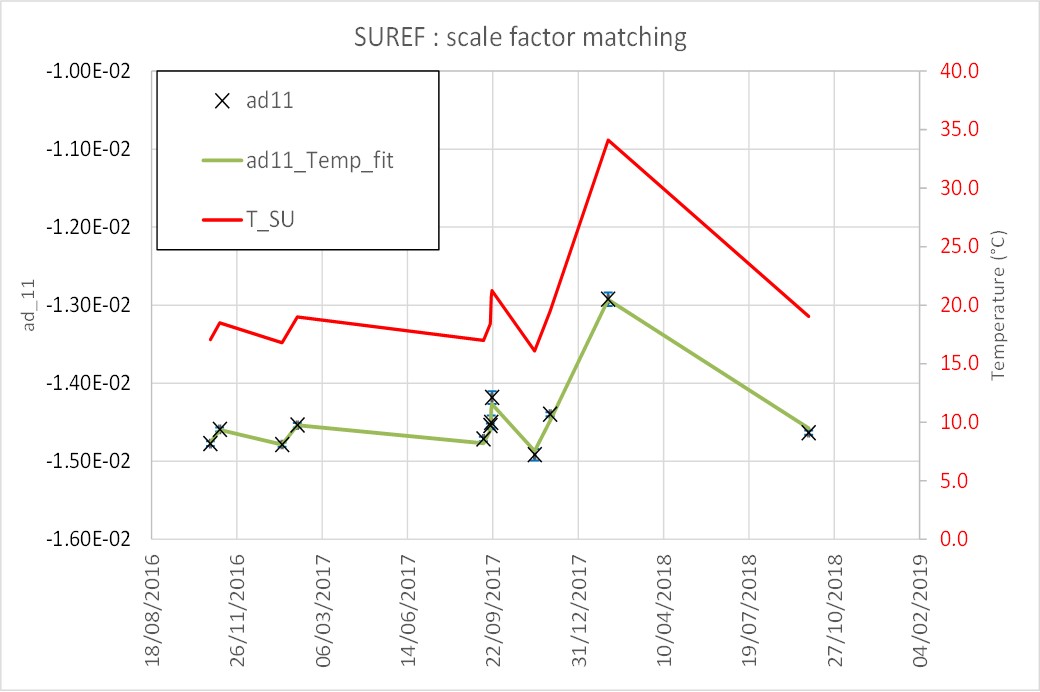}
\includegraphics[width=0.5 \textwidth]{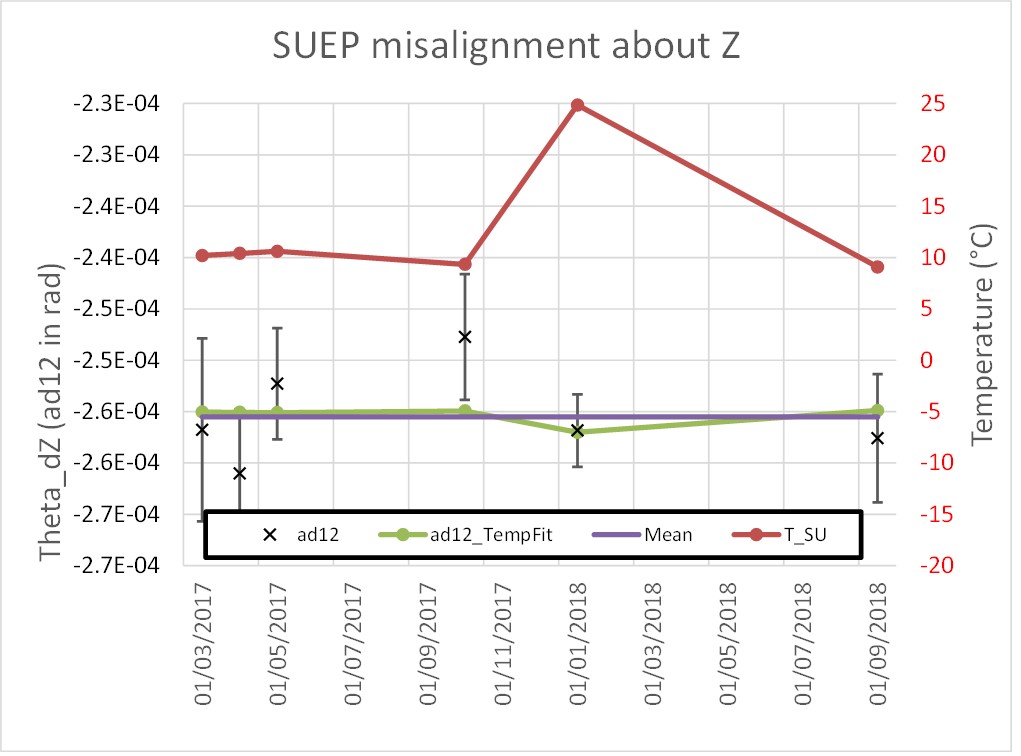}
\includegraphics[width=0.5 \textwidth]{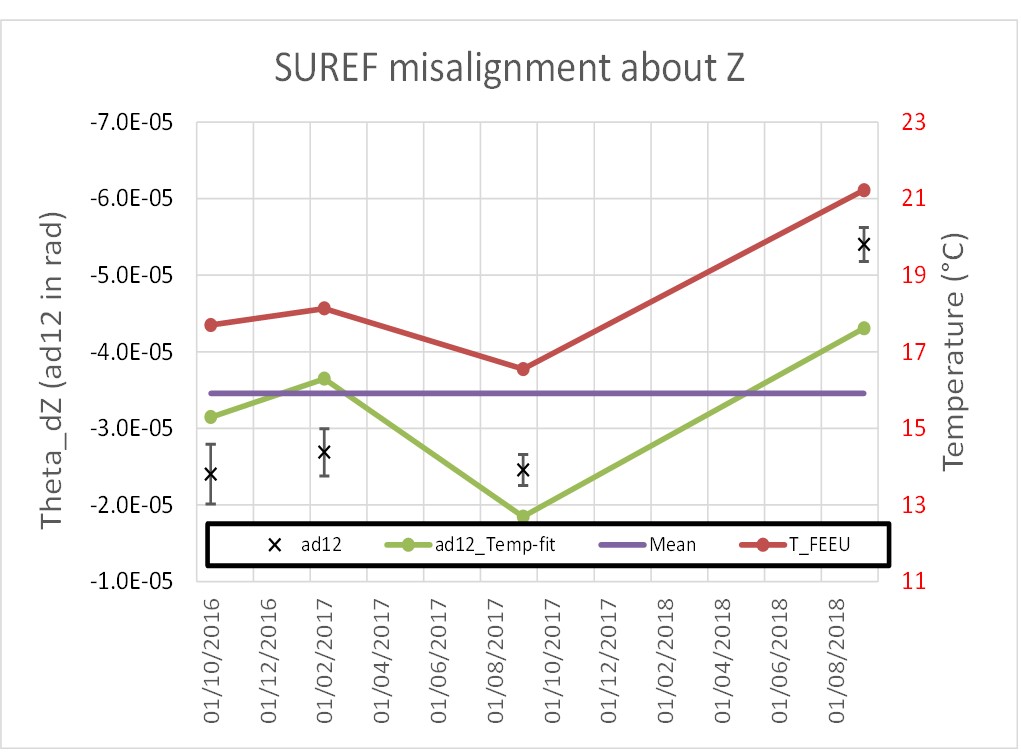}
\includegraphics[width=0.5 \textwidth]{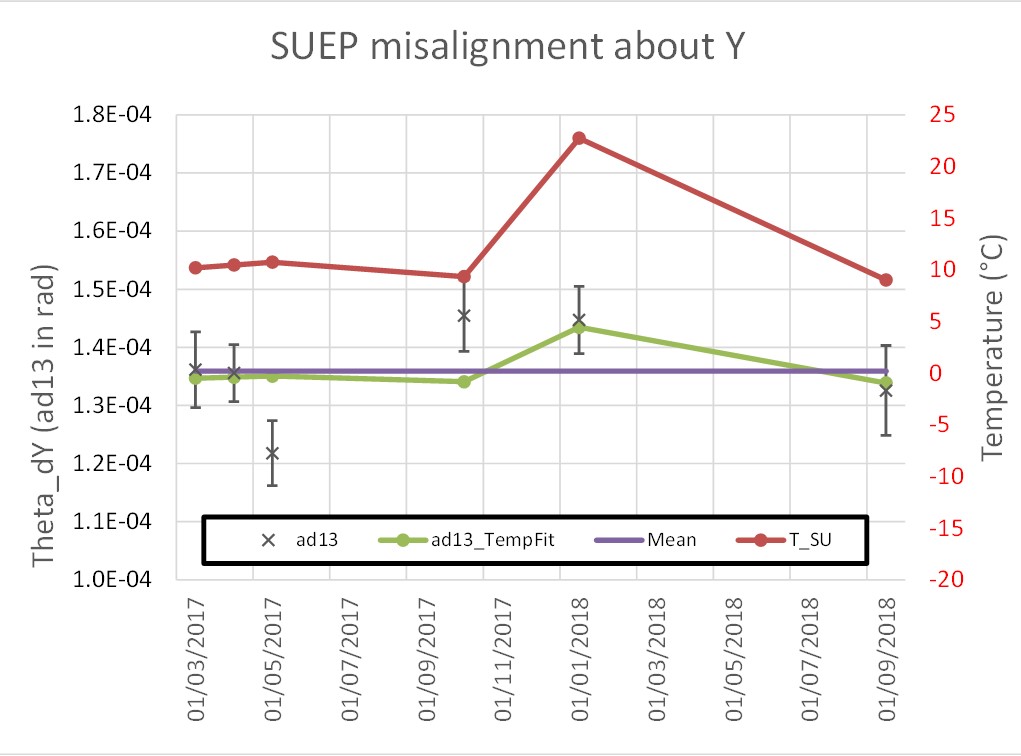}
\includegraphics[width=0.5 \textwidth]{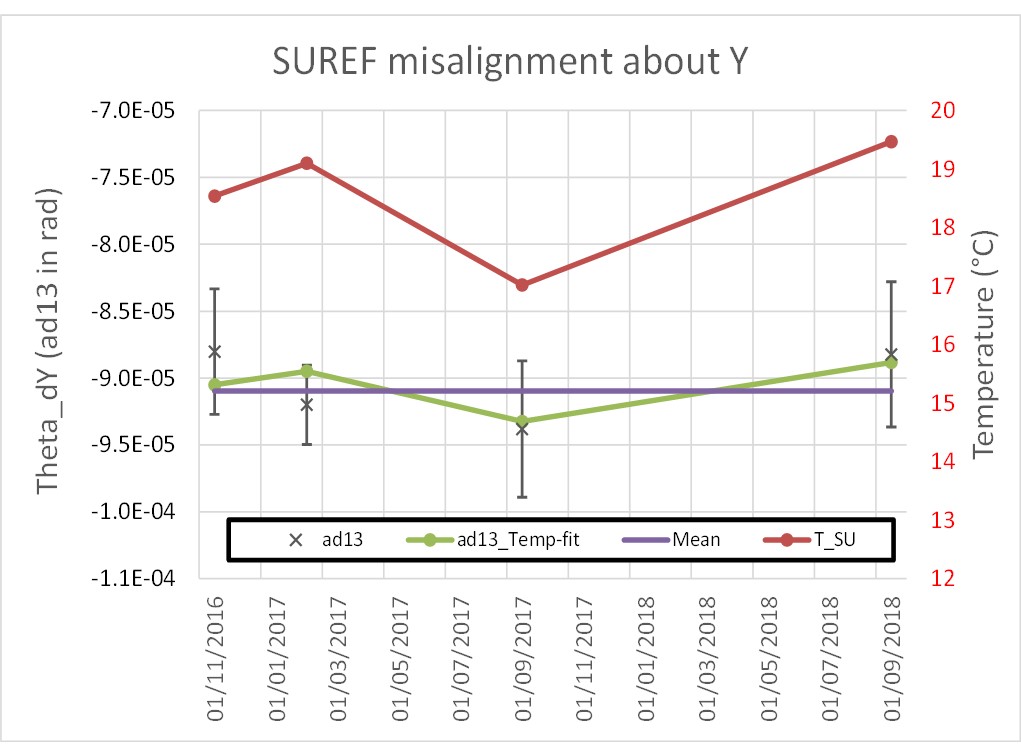}
\caption{In-orbit estimation of elements $a_{d1i}$ for SUEP and SUREF. The red line shows the mean temperature whereas the green line (Temp fit) represents the modeling of the considered parameter as a linear fit to the temperature. All estimates are plotted with their uncertainties even though they are small and not apparent.}
\label{fig_ad11} 
\end{figure*}

For calibration purposes, a much significant common acceleration is imposed to the satellite to amplify the effects of these parameters. A sine linear acceleration at frequency $f_{\rm cal}$ is applied to the satellite: along the $X$ axis to estimate $a_{d11}$, along the $Y$ axis for $a_{d12}$ and along the $Z$ axis for $a_{d13}$. The sine amplitude of the common acceleration is set to $5\times{}10^{-8}$m\,s$^{-2}$ by biasing the output of the accelerometer used as reference by the DFACS. The $a_{d1m}$ parameter is estimated at $f_{\rm cal}$ as the ratio of the differential measurement over the common mode acceleration applied to the satellite.
The resulting estimations of $\left[a_{d11}, a_{d12}, a_{d13} \right]$ for the two sensors are shown in \Fref{fig_ad11}. 

\begin{figure*}
\includegraphics[width=0.5 \textwidth]{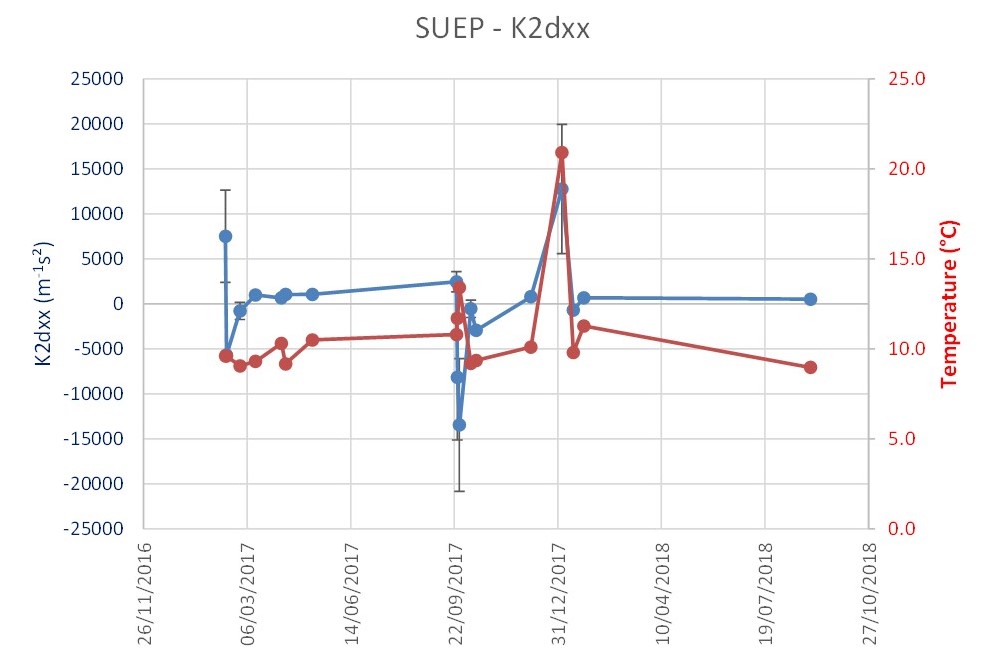}
\includegraphics[width=0.5 \textwidth]{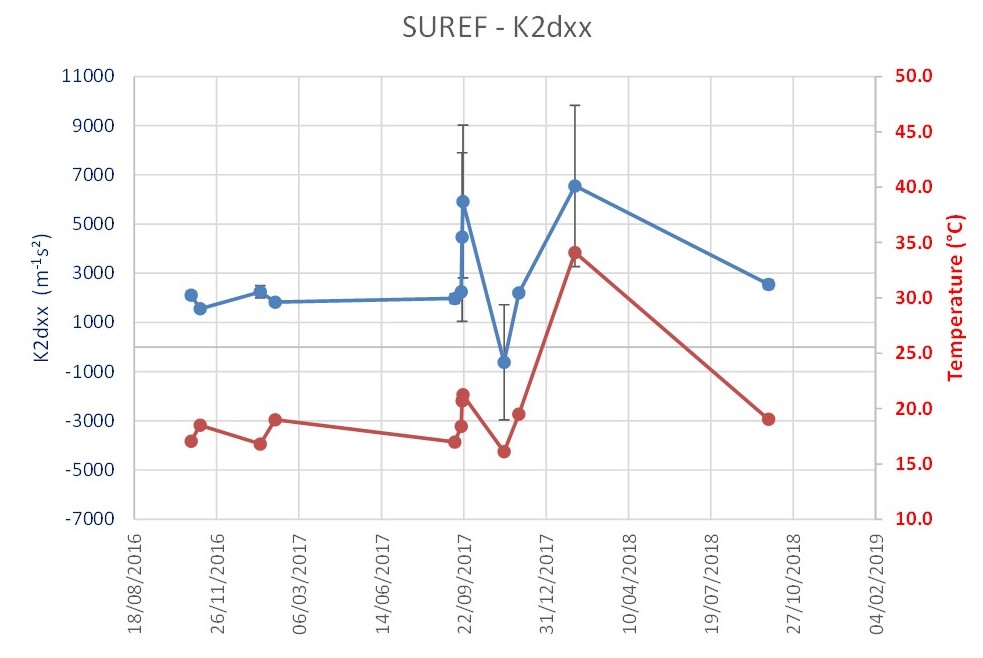}
\caption{In-orbit estimation of the differential quadratic factor $K_{2d,xx}$ for SUEP and SUREF (blue curves), and mean temperature (red line) during the calibration session.}
\label{fig_K2dxx} 
\end{figure*}

From the calibration session dedicated to $a_{d11}$, the quadratic term $K_{2d,xx}$ is also estimated at $2f_{\rm cal}$ as two times the ratio of the differential measurement over the square of the common mode $(5\times{}10^{-8}$m\,s$^{-2})^2$. As shown in \Fref{fig_K2dxx}, a sensitivity of the quadratic term to the temperature seems to be visible but the uncertainties of their estimation increasing also with temperature variations may moderate this observation. It was also observed that the sessions given at the highest temperatures in the figure were performed during phases of temperature increase or decrease and thus non-stationary conditions, contributing to increase the estimation error. However, the EP sessions were performed at better temperature conditions and thus with limited impact of quadratic term potential fluctuations.

The mean values of the calibration parameters have been estimated over all sessions and are presented in \Tref{tab_ad11} for both SU. The temperature sensitivity is only presented for $a_{d11}$ parameter because too few calibration sessions were performed for the others parameters with different temperatures. The $a_{d11}$ sessions were performed at different temperatures but without favouring the temperature variations of the SU or the FEEU on the contrary to the acceleration bias thermal sensitivity in \Sref{therm_sys}. Therefore it is difficult to decorrelate the two sources of temperature variation and only an upper bound is given.

\begin{table}
\caption{{\label{tab_ad11}}Mean values of the calibrated instrument parameters ($\pm{}$ standard deviation over all estimations at about +18\textdegree{}C for SUREF and about +10\textdegree{}C for SUEP). An upper bound estimation of the thermal sensitivity to SU temperature ($T_{SU}$) and to FEEU temperature ($T_{\rm FEEU}$) is given at the two last lines of the table.}
\begin{tabular}{lccl} 
\textbf{Parameter} & \textbf{SUREF} & \textbf{SUEP}      & \textbf{Unit}  \\ 
\hline
$a_{d11}$ & $(-1.46\pm{}0.02)\times{}10^{-2}$ & $(+0.85\pm{}0.02)\times{}10^{-2}$& \\
$a_{d12}$ & $(-0.35\pm{}0.15)\times{}10^{-4}$ & $(-2.56\pm{}0.05)\times{}10^{-4}$ & rad \\
$a_{d13}$ & $(-0.91\pm{}0.03)\times{}10^{-4}$ & $(+1.36\pm{}0.09)\times{}10^{-4}$ & rad \\
$K_{2d,xx}$ & $2409\,(\pm{}1650)$ & $-1037\,(\pm{}4800)$ &  m$^{-1}$\,s$^{2}$ \\
$\partial a_{d11} / \partial T_{\rm SU}$ & $1.3\times{}10^{-4}$ & $2.3\times{}10^{-4}$ & \textdegree{}C $^{-1}$\\
$\partial a_{d11} / \partial T_{\rm FEEU}$ & $1.8\times{}10^{-5}$ & $9\times{}10^{-5}$ & \textdegree{}C $^{-1}$ \\
\end{tabular}
\end{table}

\begin{figure}
\includegraphics[width=0.95 \textwidth]{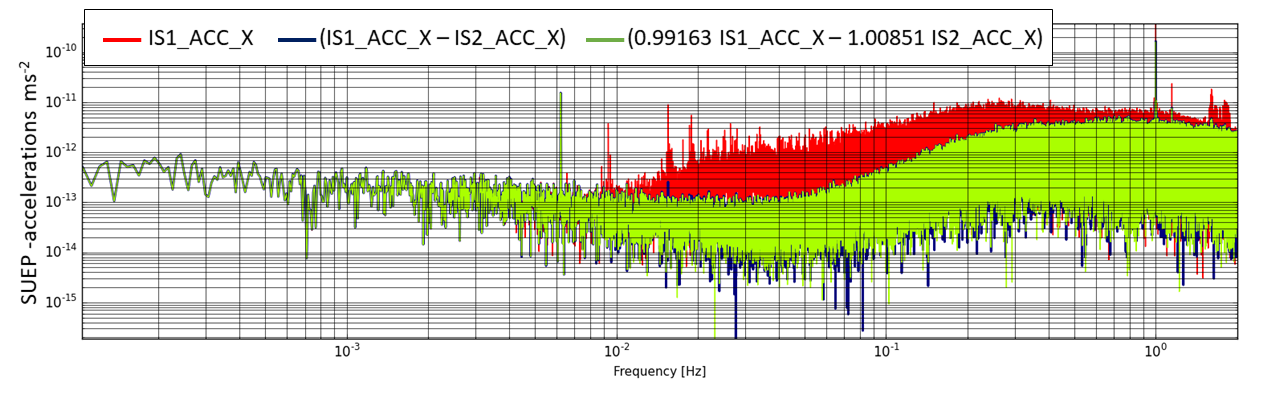}
\caption{DFT modulus of the measured accelerations of SUEP in session\#218. DFACS is controlled according to the IS2 output which is almost canceled. In red, the inner test-mass acceleration; in blue, the differential acceleration without calibration; in green, differential acceleration with calibrated parameters.}
\label{fig_suep_cor} 
\end{figure}

In order to illustrate the effect of the calibration correction process, \Fref{fig_suep_cor} shows the modulus of the Discrete Fourier Transform (DFT) of the difference of acceleration with and without calibration. Without calibration, the instrument scale factors are known with a few percent accuracy allowing the differential measurement to reject most of the common signal in particular the signal outside of the DFACS bandwidth of 0.01Hz. When the scale factors are adjusted with $a_{d11}$ set to $8.44\times{}10^{-3}$ (determination realised with the calibration session the closest to the plotted session \#218), the calibrated differential acceleration is improved in particular between 0.01\,Hz and 0.02\,Hz as shown in \Fref{fig_suep_cor}. The noise is also a little bit improved and tests performed with and without calibration show a 10\% improvement of the accuracy of the EP test.

%
%
%

%
%
%
%

\section{Thermal analysis} \label{therm_sys}

In Refs. \cite{touboul17,touboul19} the temperature sensitivity of the accelerometers appears as the main contributor to the systematic error budget. This contributor comes mainly from two sources: the time variation of temperature in the sensor unit, $\delta T_{\rm SU}$, and the time variation of the temperature in the FEEU, $\delta T_{\rm FEEU}$. In \Eref{eq_xacc}, the two biases terms $B_{0x}^{(d)}$ and $b_{1x}^{(d)}$ contain time variations due to thermal effects. Since it is difficult to separate the contribution of $\delta T_{\rm SU}$ and of $\delta T_{\rm FEEU}$ to each bias, we gather the effects in a global term $\Gamma^{(d)}_{\rm th}$ which can be modelled as:
\begin{equation}  \label{eq_therm}
\Gamma^{(d)}_{\rm th}=\lambda_{\rm SU}\delta T_{\rm SU}+\lambda_{\rm FEEU}\delta T_{\rm FEEU}.
\end{equation}

The acceleration measurement temperature sensitivities to the SU $\lambda_{\rm SU}$ and to the FEEU $\lambda_{\rm FEEU}$ are obtained with dedicated sessions where a stimulus at SU level or at FEEU level is applied and the acceleration is measured (\Sref{therm_calib}). The temperature is continuously measured during science sessions in different parts of the payload case and of the satellite in order to estimate the systematic effect. In Refs. \cite{touboul17,touboul19}, the variation of temperature at $f_{\rm EP}$ does not emerge from the temperature probe noise. This noise integrated over several orbits was taken into account as the upper-bound limit of the systematic temperature variation at $f_{\rm EP}$. In this section we better estimated the thermal sensitivities and temperature variations that lead in \Sref{sum_sys} to a final budget of systematic errors.

	\subsection{Thermal sensitivity characterisation} \label{therm_calib}
	
		\subsubsection{Heating resistors and thermal sensors.}
		
\begin{figure} [h]
	\centering
	\includegraphics[scale=0.29]{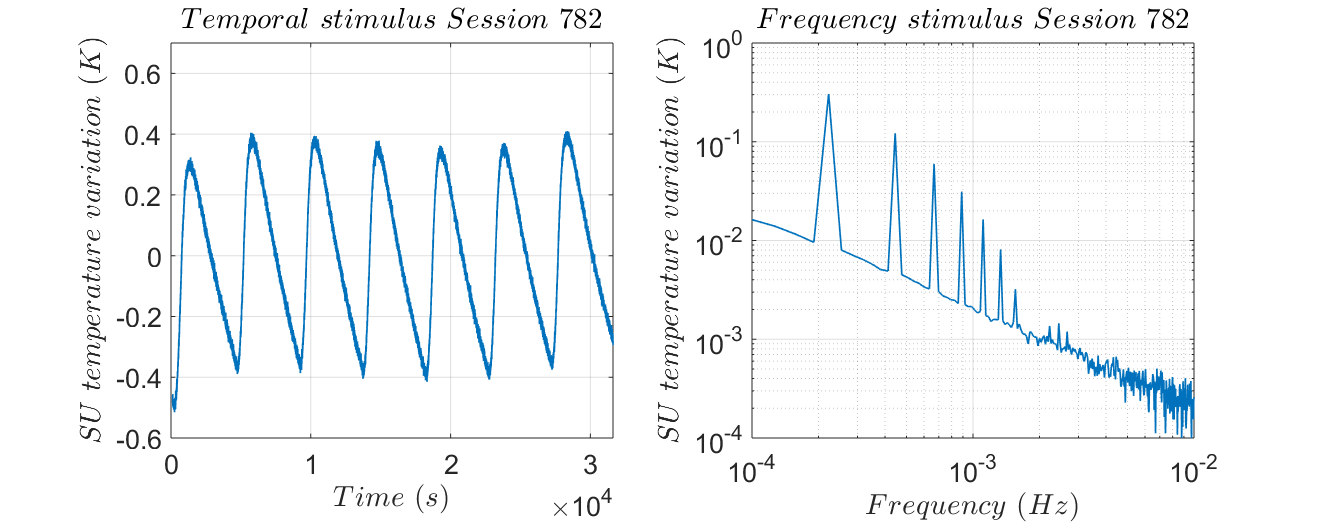}
	\caption{Temperature stimulus as a function of time (left) and its DFT modulus (right) for session 782.}
	\label{stimulus}
\end{figure}

In order to characterise the thermal behaviour of the instrument, a temperature variation is applied to the SU or to the FEEU at a stimulus frequency $f_{\rm sti}$ close to the frequencies of interest ($f_{\rm orb}$, $f_{\rm EP}$) similar to the example in \Fref{stimulus}.

\begin{figure}
	\centering
	\includegraphics[scale=0.45]{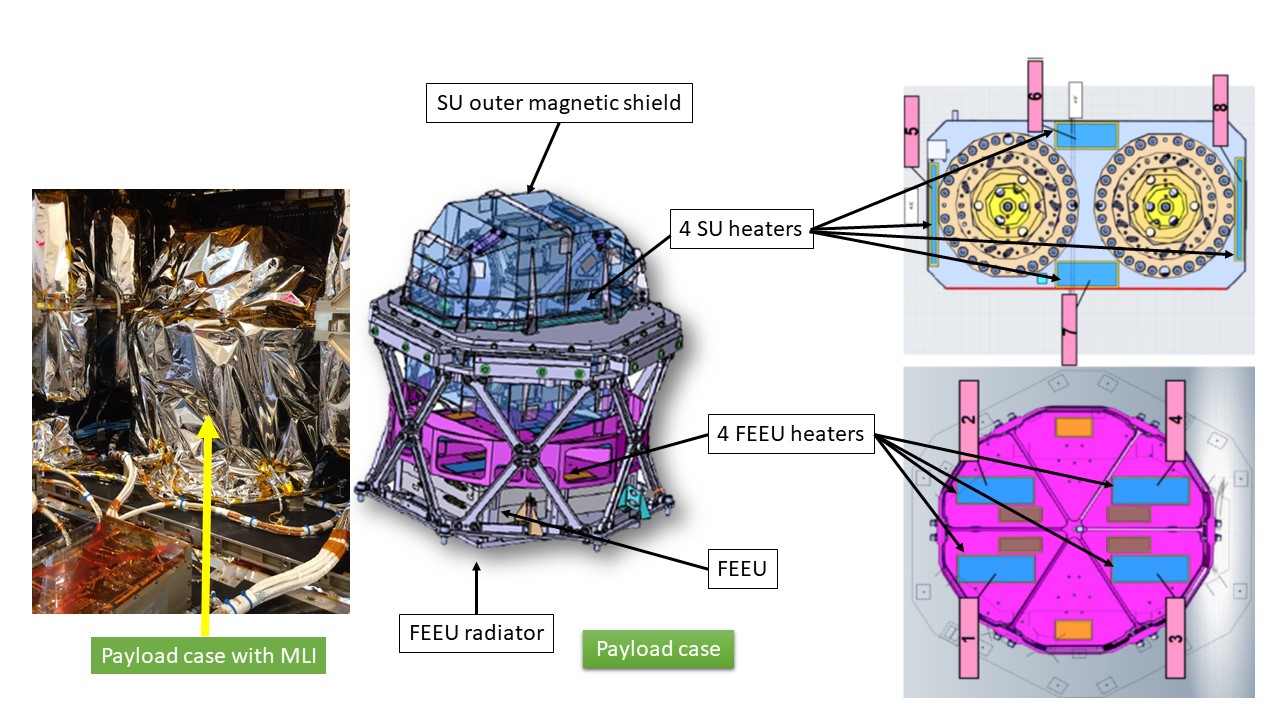}
	\includegraphics[scale=0.46]{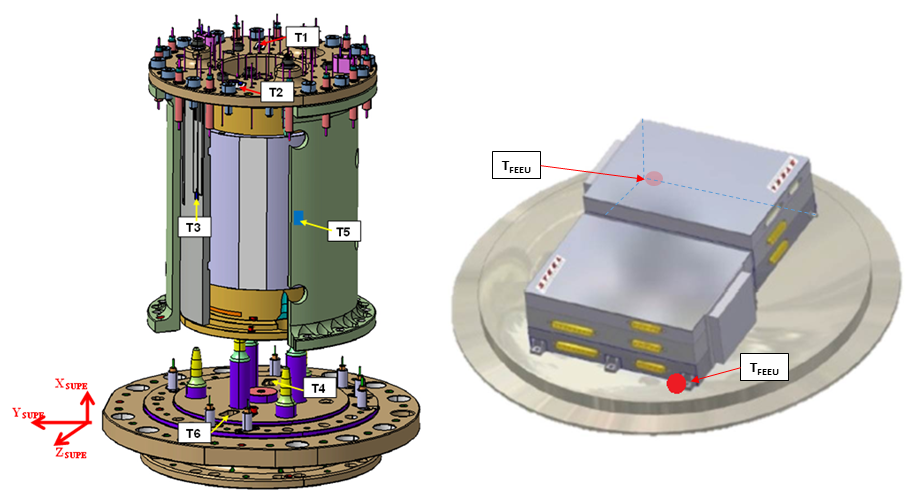}
	\caption{Upper panel: on left, payload case covered with Multi-Layer Insulation (MLI); on right, heaters 1 to 4 are placed on the FEEU plate and heaters 5 to 8 are placed on the SU plate called SU mechanical interface (SUMI). Lower panel : temperature probes in the SU (left) and at FEEU interface (right).}
	\label{probes}
\end{figure}

To do so, eight double-layer heaters were placed near the SU and near the FEEU (\Fref{probes}). These heaters are used to generate the temperature stimulus with a current compensation to minimise the induced magnetic field. Pt1000 temperature probes are located inside each SU core and at the interface of each FEEU. In the SU the temperature probes are numbered from T1 to T6 (\Fref{probes}). The mean of T6 and T4 is used to evaluate the SU temperature $T_{\rm SU}$. Five probes are integrated in each FEEU whose temperature $T_{\rm FEEU}$ is evaluated with the one at the interface, close to the Temperature Reference Point of the unit.

		\subsubsection{Thermal sessions for the characterisation of the instrument.}
		
As shown in \Fref{stimulus}, the heaters on the SU plate, respectively the FEEU plate, are switched on during a time $T_{\rm p}$ with a period $T_{\rm sti}=1/f_{\rm sti}$. 
Several constraints have determined the choice of $f_{\rm sti}$.
The value of $f_{\rm sti}$ has been defined as a non-multiple of $f_{\rm orb}$ in order to avoid a combination with the natural orbital frequency while applying the temperature stimulus. 
The shorter the heating time $T_{\rm p}$, the lower is the temperature variation amplitude because not much heating power is injected on the system. Consequently, the period $T_{\rm sti}$ is constrained by the heating period and thus higher frequency of stimulus is difficult to obtain.
A longer heating period improves the signal to noise ratio, but also induces an increase of the mean temperature (i.e. a drift) as the satellite cannot dissipate at the same rate the amount of heat power created by the stimulus. The characterisation was performed in a limited range of temperature of a few degrees in order to prevent additional drifts. The number of cycles, defined by $\frac{T_0}{T_{\rm sti}}$, $T_0$ being the duration of the thermic session, has to be as high as possible to better characterise the thermal sensitivity but has also to be constrained to limit the temperature drift.

The list and durations of the thermal sessions are presented in \Tref{list}.

\begin{table}
\caption{{\label{list}} List of dedicated sessions to instrument temperature sensitivity and their durations.}
\begin{tabular}{lllll}
\textbf{Session} & \textbf{Part of the instrument tested} & \textbf{$T_0$} & \textbf{$T_{\rm sti}$(s)} & \textbf{$T_{\rm p}$(s)} \\
\br
\textbf{266}     & SUEP FEEU     & 8h00m00s       & 1500               & 300                \\
\textbf{270}     & SUEP SU       & 8h00m00s       & 4500               & 500                \\
\textbf{298}     & SUREF FEEU    & 4h00m45s       & 321                & 64                 \\
\textbf{300}     & SUREF SU      & 4h00m45s       & 321                & 64                 \\
\textbf{304}     & SUREF FEEU    & 4h12m28s       & 1082               & 200                \\
\textbf{306}     & SUREF SU    & 4h12m28s       & 1082               & 120                \\
\textbf{314}     & SUEP FEEU     & 4h12m28s       & 1082               & 200                \\
\textbf{316}     & SUEP SU       & 4h12m28s       & 1082               & 120                \\
\textbf{320}     & SUEP FEEU     & 4h00m45s       & 321                & 64                 \\
\textbf{322}     & SUEP SU       & 4h00m45s       & 321                & 64                 \\
\textbf{758}     & SUEP FEEU     & 8h00m00s       & 321                & 128                \\
\textbf{760}     & SUEP FEEU     & 8h00m00s       & 1082               & 432                \\
\textbf{782}     & SUREF SU      & 8h00m00s       & 4500               & 500               
\end{tabular}
\end{table}

	\subsubsection{Estimation of the temperature sensitivity}

\begin{figure}
    \centering
 	\includegraphics[scale=0.35]{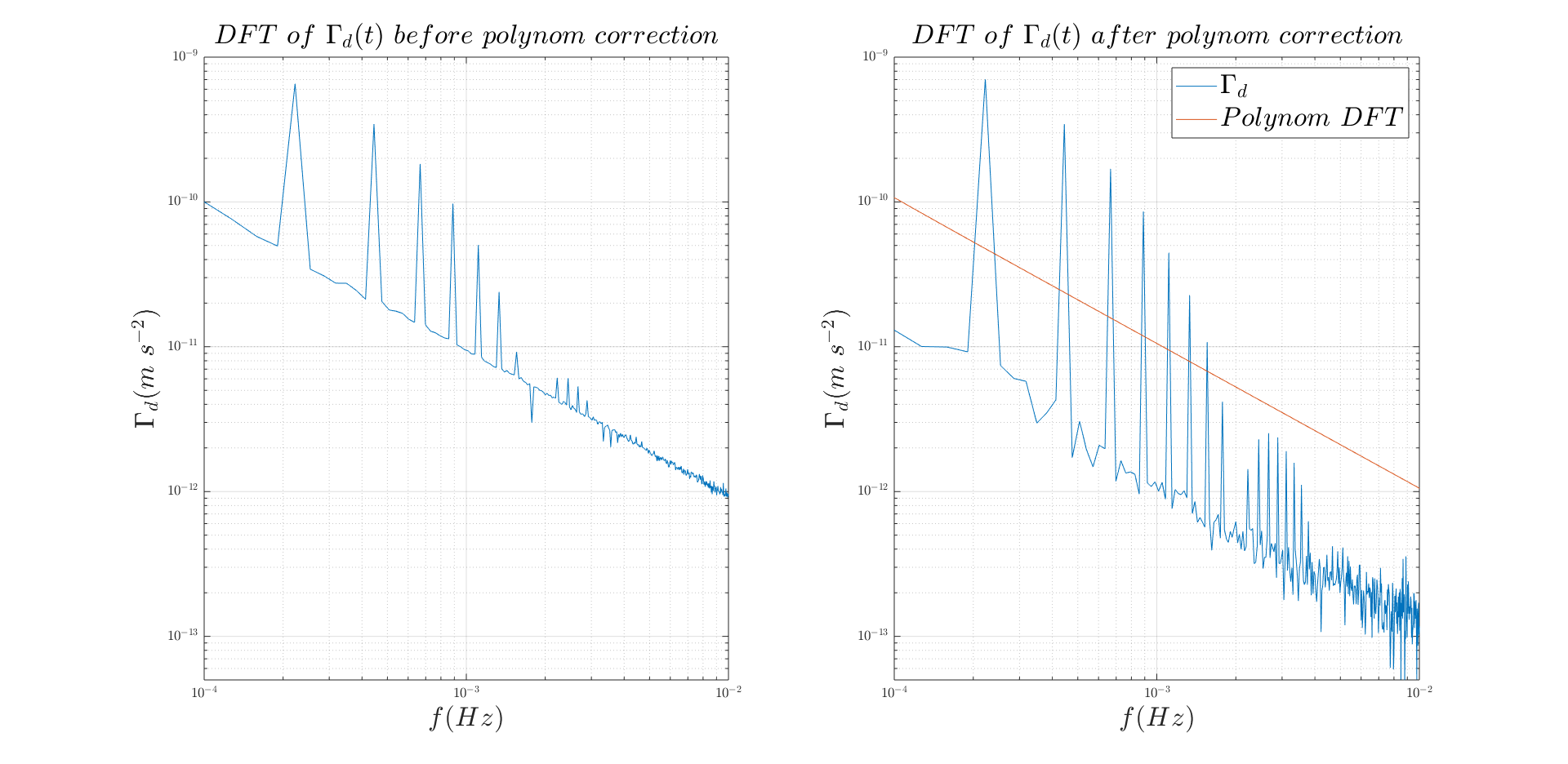}
 	\caption{DFT of the differential acceleration in session 782 without (left) and with correction (right) of the drift by a second order polynomial. The DFT of the polynomial is also plotted on right panel. The differential acceleration data is corrected from the Earth's gravity gradient effect in both figures.}
 	\label{fig_poly}
 \end{figure}

The process to estimate the sensitivity of the measured differential acceleration to temperature variations of the FEEU and of the SU is detailed in Ref. \cite{dhuicque21}. Here we report only the outline and the results. In order to estimate the sensitivities  $\lambda_{\rm SU}$ and $\lambda_{\rm FEEU}$ given the measured values of $\Gamma^{(d)}_{\rm th}$, $\delta T_{\rm SU}$ and $\delta T_{\rm FEEU}$ (\Eref{eq_therm}) we apply the following preprocessing: 
\begin{itemize}
\item Since the measurements can be affected by slightly different scale factors on each test mass, the measured differential acceleration must be corrected from the small projection of the common mode acceleration \cite{rodriguescqg1}. This correction is proportional to the  differential scale factor which is determined thanks to dedicated calibration sessions (see \Sref{calib}). Note that during thermal sessions, the satellite drag-free system is not activated, so the common mode acceleration includes the satellite drag due to the atmosphere which is mainly visible at the orbital frequency $f_{\rm orb}$. To break free from this effect a scale factor is applied in order to correct the differential acceleration. 
\item Some data are missing (1 over 100 000 data samples): to cope with missing data, the average of the nearest 10 points is taken which is sufficient here for thermal analysis (but not for the EP test as explained in Ref. \cite{baghi15}).
\item We can distinguish two parts of the temperature and acceleration signals: the long-term part (drift) and the short-term part in which the stimulus appears. In order to get rid of the long-term part, a polynomial fit of order 2 is subtracted to absorb the signal drift. \Fref{fig_poly} illustrates the impact of the drift correction in the DFT.
\end{itemize}

\begin{figure}
	\centering
	\includegraphics[scale=0.35]{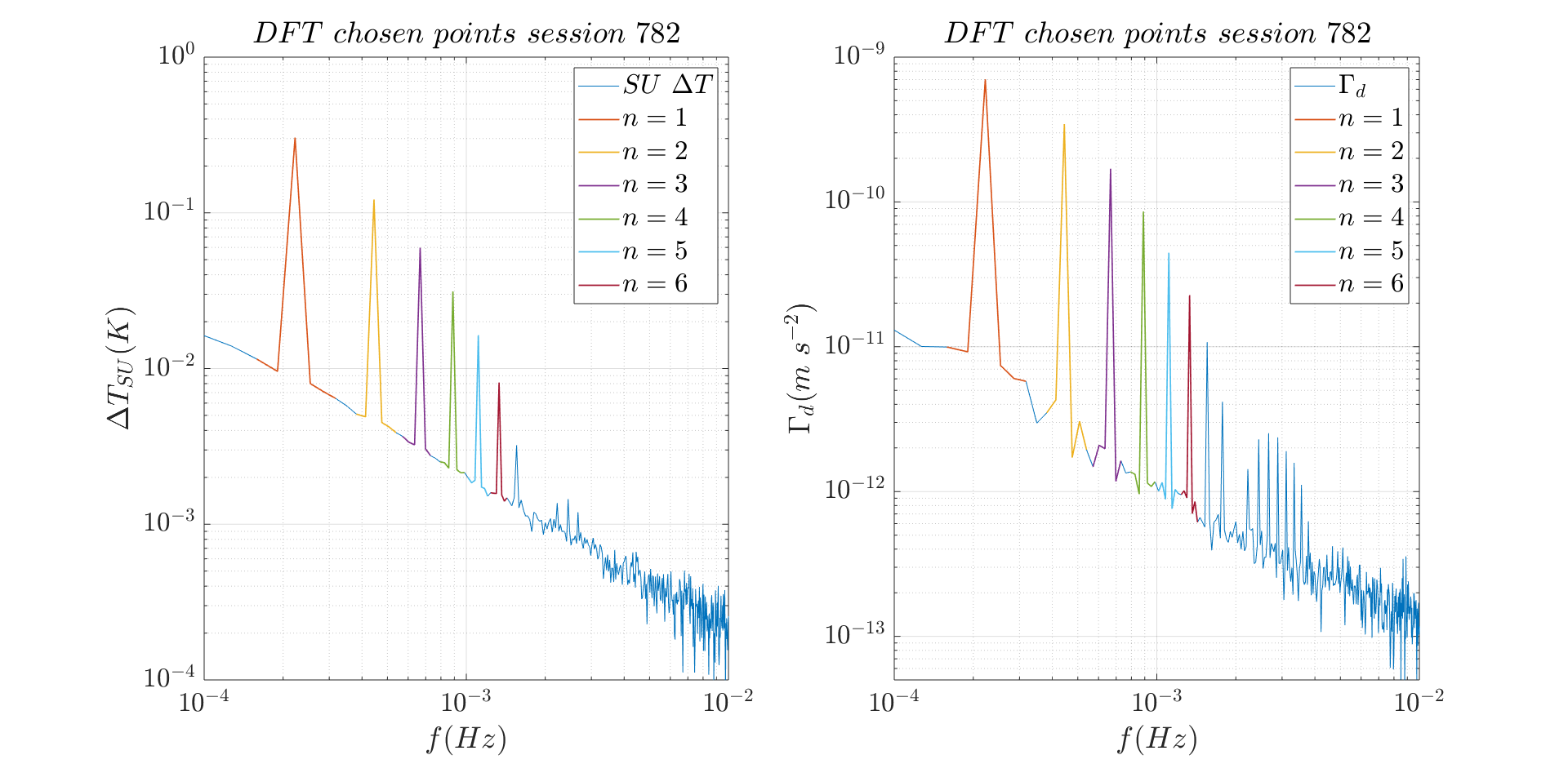}
	\caption{Discrete Fourier transform of the temperature variation (left) and  of the differential acceleration (right) for session 782. The effect of Earth's gravity gradient has been corrected in the differential acceleration data. The points taken into account around each harmonic $n f_{sti}$ are of the same colour.}
	\label{freqmeth}
\end{figure}

The sensitivities are estimated using a least-square algorithm in the frequency domain. A DFT is applied to the measurements of the differential acceleration and of the temperature data, converting the measurement equations in the time domain into the same number of measurement equations in the Fourier domain.  Since we choose to analyse segments containing an integer number of periods $T_{\rm sti}$, the frequency $f_{\rm sti}$ and its harmonics $n f_{\rm sti}$ correspond exactly to discrete frequencies of the DFT. The least-square regression is then applied to small frequency bands around $n f_{\rm sti}$. 
\Fref{freqmeth} shows an example of DFT modulus and the samples around the stimulus frequencies taken into consideration for the least-square regression.

Ref. \cite{dhuicque21} gives more details on this method and presents also an other method operating directly in the time domain. Both methods lead to comparable results. \Tref{thresults} reports the results at two particular frequencies of interest $f_{\rm EP2}=9.24 \times{}10^{-4}$\,Hz and $f_{\rm EP3}=3.11 \times{}10^{-3}$\,Hz which are the two frequencies used during EP science sessions. Note that SUEP shows a frequency dependence more obvious than SUREF (see Ref. \cite{dhuicque21} for more details).

%
\begin{table}
\caption{{\label{thresults}} Thermal sensitivity determination for SUREF and SUEP. The associated error combines the accuracy of the least square method and the dispersion of all tests.}
\centering
\begin{tabular}{lll} 
      & $\lambda_U$ (m\,s$^{-2}$\,\textdegree{}C$^{-1})$        \\ \hline
\textbf{SUEP: $\lambda_{\rm SU}$  at $f_{\rm EP2}$} & $(1.4  \pm{}0.15) \times{}10^{-8}$ \\
\textbf{SUEP: $\lambda_{\rm SU}$  at $f_{\rm EP3}$} & $(6.4 \pm{}0.7) \times{}10^{-9}$ \\
\textbf{SUEP: $\lambda_{\rm FEEU}$  at $f_{\rm EP2}$}       & $(7.3  \pm{}0.75) \times{}10^{-11}$  \\
\textbf{SUEP: $\lambda_{\rm FEEU}$  at $f_{\rm EP3}$}       & $(5.5 \pm{}0.5) \times{}10^{-11}$ \\
\textbf{SUREF: $\lambda_{\rm SU}$ at $f_{\rm EP}$}  & $(5.2 \pm{}2.3) \times{}10^{-9}$ \\ 
\textbf{SUREF: $\lambda_{\rm FEEU}$ at $f_{\rm EP}$}      & $(7.1 \pm{}1.5) \times{}10^{-11}$
\end{tabular}

\end{table}

	\subsection{Temperature variations at $f_{\rm EP}$} \label{therm_design} 
\subsubsection{Thermal design.}
	
\begin{figure}
	\centering
	\includegraphics[scale=0.6]{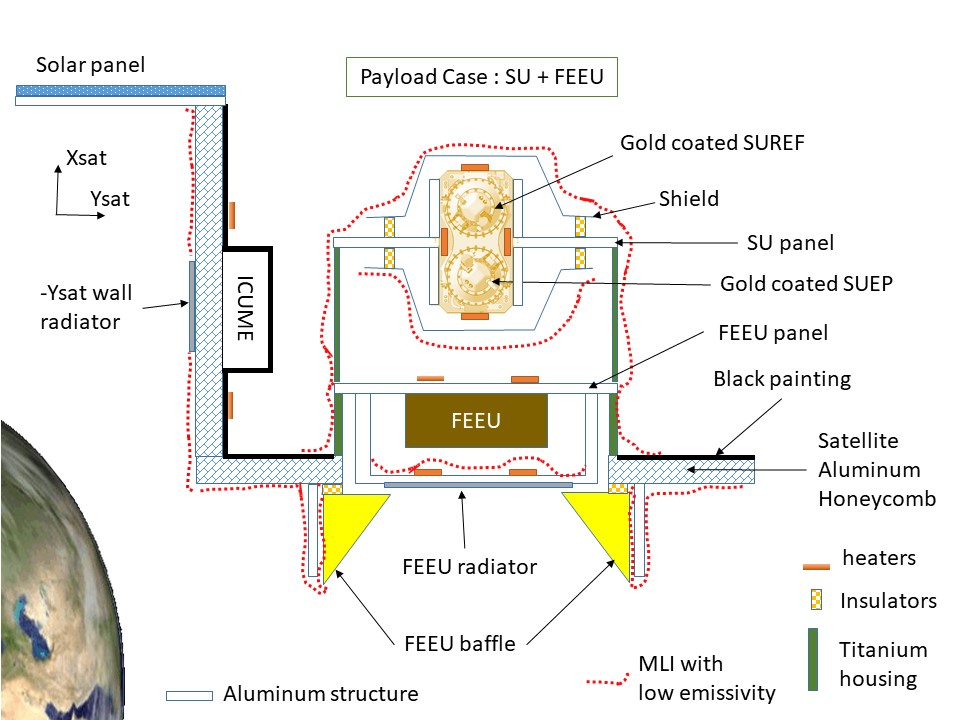}
	\caption{Satellite cut-away view with the thermal design and heater localisations.}
	\label{sat}
\end{figure}

The thermal design of the satellite was a major driver for the definition of the payload integration. In order to minimise any thermoelastic effect due to active thermal control by heaters, the use of heaters during science phases was forbidden. Although the SU does not contain any power dissipative elements, each FEEU dissipates about 3\,W and  this energy must be evacuated by a dedicated radiator (\Fref{probes}, \Fref{sat}). In order to ensure the passive thermal control of the payload case in the range of [10\textdegree{}C - 40\textdegree{}C] and of all satellite sub-equipment, some rules of design and choice of materials have been performed:

\begin{itemize}
\item the coatings of the inner satellite structure and of some equipment are derived from the Myriad line,
\item the satellite outer coating is made of Multi Layer Insulator (MLI) instead of Beta cloth to minimise the mechanical cracking and electrical discharging,
\item a reinforced MLI is used around the payload case,
\item the harnesses between the FEEU and the SU have a low thermal conductivity and are covered by MLI,
\item all inner faces of the satellite walls and the electronics surfaces are painted in black to homogenise the thermal emissivity,
\item a thermal seal is used at the interface of equipment with the satellite walls,
\item the battery is decoupled from the wall to reduce thermal conductions,
\item radiators are placed symmetrically on both walls along $Y_{\rm SAT}$ and their area has been optimised after ground tests performed in 2009 with thermal mockups representative of the satellite and of the payload case,
\item heaters are used in operation to bring the equipment to the proper temperature range for their switch on and to maintain the temperature when the equipment is switched off. A particular line of heaters is placed in the payload case, with current compensation to minimise the induced magnetic field, and exclusively used during the thermal sessions.
\end{itemize}

The thermal design was first validated during a ground campaign in 2009 involving the payload and the satellite staff with a detailed representative model of the satellite and of the payload. It allowed us to adjust the thermal model and to confirm the sizing of the conductive and the insulated elements, in particular the FEEU radiator. The FEEU radiator was identified as the key factor for the entry of temperature fluctuations at $f_{\rm EP}$ in the payload case. A larger radiator gives a better heat dissipation in the outer space, a smaller radiator gives lower coupling with the Earth's albedo and thus lower variation at $f_{\rm EP}$. A trade-off was then made for the worst case conditions of the mission: inertial pointing during the periods of the year with the largest temperature variations. The rotating modes of the satellite are considered less stringent to reach the temperature specifications thanks to a better homogenisation of the temperature.

	\subsubsection{In-flight characterisation of the thermal behaviour.} \label{flight_therm} 
After the commissioning phase \cite{rodriguescqg4}, the rotating mode with two frequency rates was set as the baseline for the EP test. The thermal sensitivity characterisation in \Sref{therm_calib} shows values of sensitivity higher than expected by several orders of magnitude. Fortunately, the temperature variations are also several orders of magnitude smaller in rotating mode. That leads in Ref. \cite{touboul19} to determine upper bound values based on the observation of 320 orbits and to set the systematics to about $9\times{}10^{-15}$.
In order to reduce this upper bound limit value, the thermal model was verified in flight with dedicated sessions. The objective was to confirm ground tests such that the variations of temperature are due to the Earth's albedo coming on the FEEU radiator. As shown in \Fref{sat}, a baffle is supposed to protect the radiator from Earth's thermal radiation but some grazing light can enter. The relative rotation of the satellite with respect to the Earth modulates the illumination on the FEEU radiator and thus the temperature at $f_{\rm EP}$. These variations are transmitted to the FEEU panel and eventually to the SU panel. 

Different thermal stimulus have been applied at different stage of the satellite with different types of session:
\begin {itemize}
\item 20 orbits performed with SUREF (session \#568) and with SUEP (session \#564) with a temperature variation on $Y_{\rm sat}$ satellite wall;
\item 465 cumulated orbits (session \#676-684) with increase of Earth's albedo on the radiator placed in $X_{\rm sat}$, this session was the major thermal experiment as the temperature variations were amplified and better detected;
\item a particular session (\#446-448) for aeronomy experiment where the satellite was spun at $3f_{\rm orb}$ during which both SU were switched on at the same time; 
\item sessions (\#668-670 and \#672-674) with temperature variations realised with satellite heaters near the FEEU stage.
\end {itemize}

The strategy of the tests was to eliminate the potential temperature sources which did not come from the main radiator on $X_{\rm sat}$ and then confirm that temperature variations in the SU are correlated to the FEEU ones, the latter being also correlated to the $X_{\rm sat}$ radiator ones. The last part of the strategy is to assess that SUEP and SUREF behave similarly.

\begin{figure} \center
\includegraphics[width=0.99 \textwidth]{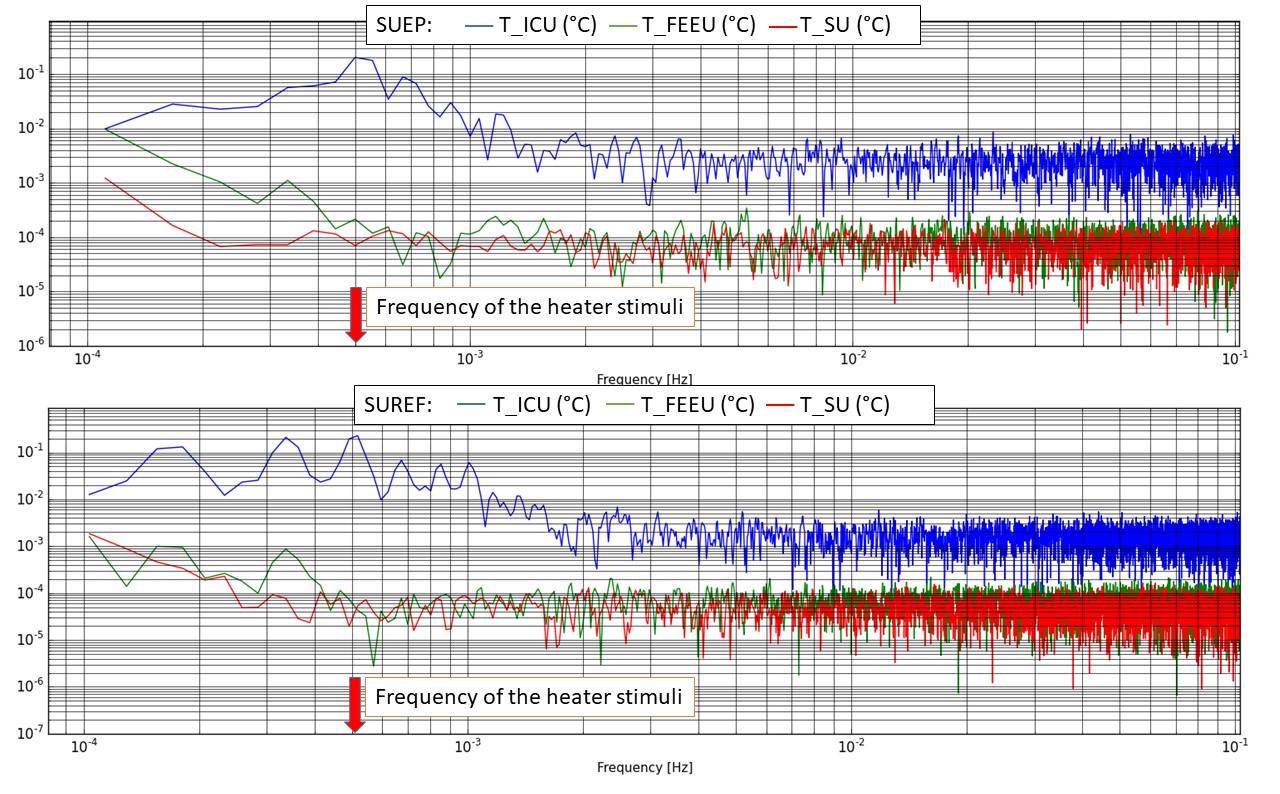}
\caption{DFT modulus of the measured temperature at FEEU (green), SU (red) and ICU (blue) interface for SUEP (upper panel, Session \#564) and for SUREF (lower panel, Session \#568).}
\label{fig_temp_icu} 
\end{figure}

The first thermal session consists in applying temperature variations on the satellite wall supporting the Interface Control Unit Mechanical Ensemble (ICUME) of the payload. Indeed, each of the two $Y_{\rm sat}$ walls supports a radiator that faces the Earth and directly undergoes the albedo at $f_{\rm EP}$. In inertial pointing a thermal stimulus was applied on one of the $Y_{\rm sat}$ wall heaters with a period of 2000\,s. Due to operational constraints, the period of the heating could not be regular (on the contrary to SU and of FEEU heaters) and thus the stimulus frequency is not accurate and can vary during the session between $4\times{}10^{-4}$\,Hz and $5\times{}10^{-4}$\,Hz. In \Fref{fig_temp_icu}, the DFT modulus of the temperature probe measurements for the FEEU (in green) and for the SU (in red) shows attenuations of the temperature stimulus by a factor 1000 or more with respect to ICU temperature (in blue) whose measurement is close to the wall heaters. Consequently, the impact of the Earth's albedo on this part of the satellite walls has been considered negligible in the final assessment.

\begin{figure} \center
\includegraphics[width=0.99 \textwidth]{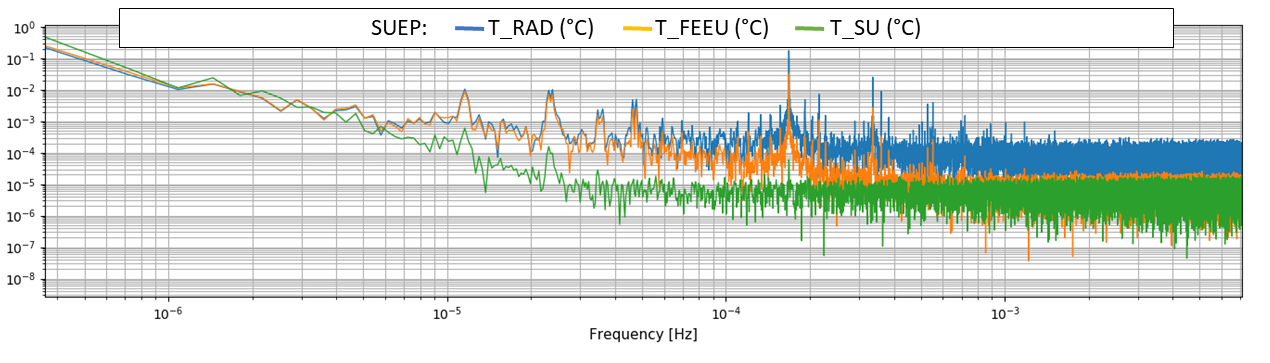}
\includegraphics[width=0.8 \textwidth]{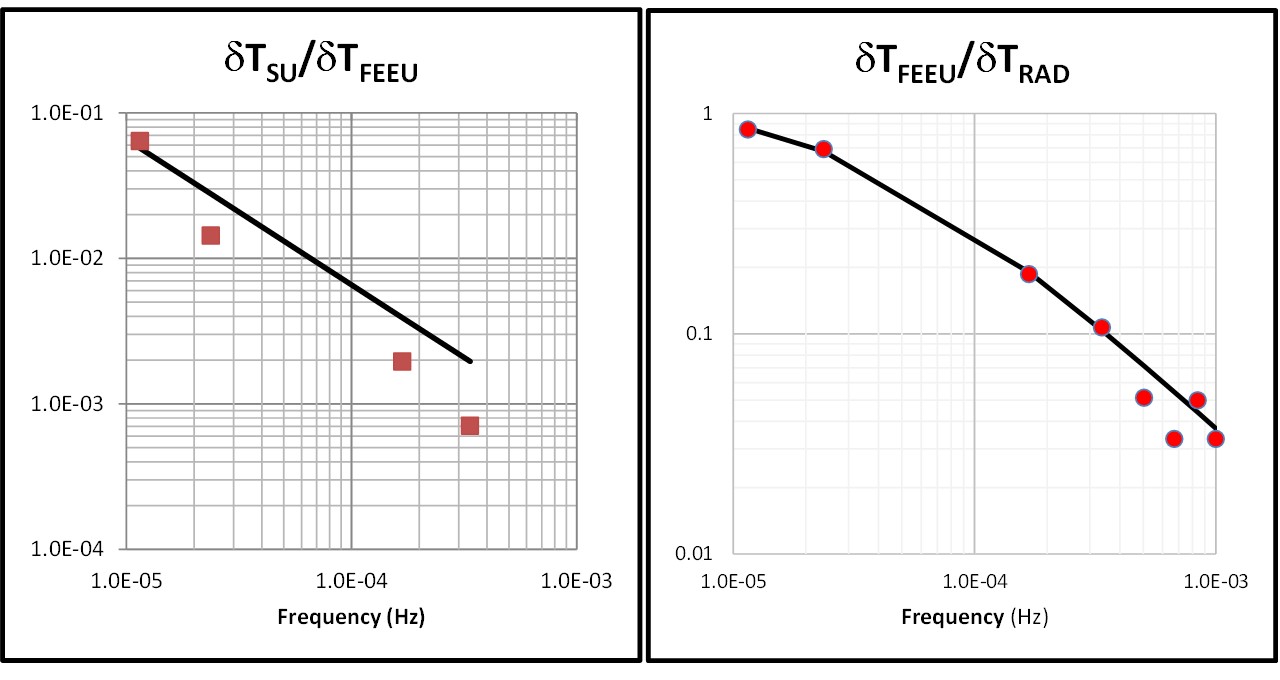}
\caption{Session \#676-684. Upper panel: DFT modulus of the measured temperature for SUEP's FEEU (orange), SUEP's SU (green) and Radiator interface (blue). Lower panel:  in red marks, ratio of the FEEU and SU temperature variations (left) and the ratio of the radiator and FEEU temperature variations (right), at different frequencies; the black line is a fit of a first order low-pass filter.}
\label{fig_temp_spicho} 
\end{figure}

The second thermal session is a very long session of 465 continuous orbits (32.3 days, session \#676-684) with the satellite tilted by $30^{\circ}$ about the spin axis in inertial pointing in order to increase the entry of albedo light into the FEEU radiator. This configuration induces an amplification of the temperature variations at the radiator level as shown in \Fref{fig_temp_spicho}. At orbital frequency $1.68\times{}10^{-4}$\,Hz and its harmonics, the FEEU and SU temperature probe measurements show a significant signal response. The temperature variations of the FEEU $\delta T_{\rm FEEU}$ are attenuated with respect to the radiator temperature variations $\delta T_{\rm RAD}$ by a frequency-dependent factor: a little bit higher than 5 at the orbital frequency and about 30 near $10^{-3}$\,Hz, i.e. close to the $f_{\rm EP}$ value in V2 mode. Frequencies higher than twice the orbital frequency have not been considered for the SU temperature variations $\delta T_{\rm SU}$ because of a response lower than the noise. Nevertheless at orbital frequency, this experiment shows an attenuation of $\delta T_{\rm SU}$ with respect to $\delta T_{\rm FEEU}$ by a factor 500. Both ratios $\delta T_{\rm SU}/\delta T_{\rm FEEU}$ and $\delta T_{\rm FEEU}/\delta T_{\rm RAD}$ vary with frequency as a first order low-pass filter.

In the third case of session (\#446-448), the two SUs were operating at the same time with the possibility to compare the thermal response with the same inputs. This session demonstrated that both SUs have the same thermal filtering behaviour and that the attenuation of the radiator temperature variation with respect to the FEEU is of 7 at $f_{\rm orb}$ and 26 at $3f_{\rm orb}$ for both instruments. These figures can be compared to the one obtained with the SUEP only in session \#676-684: in this configuration an attenuation factor has been evaluated to 6 at $f_{\rm orb}$ and 20 at $3f_{\rm orb}$. This shows that SUEP and SUREF have quite the same temperature filtering functions and that SUEP may be less filtered than SUREF. Consequently all the results of the thermal sessions performed on SUEP, much more numerous than on SUREF, can be also applied to SUREF as conservative results.

\begin{figure} \center
\includegraphics[width=0.8 \textwidth]{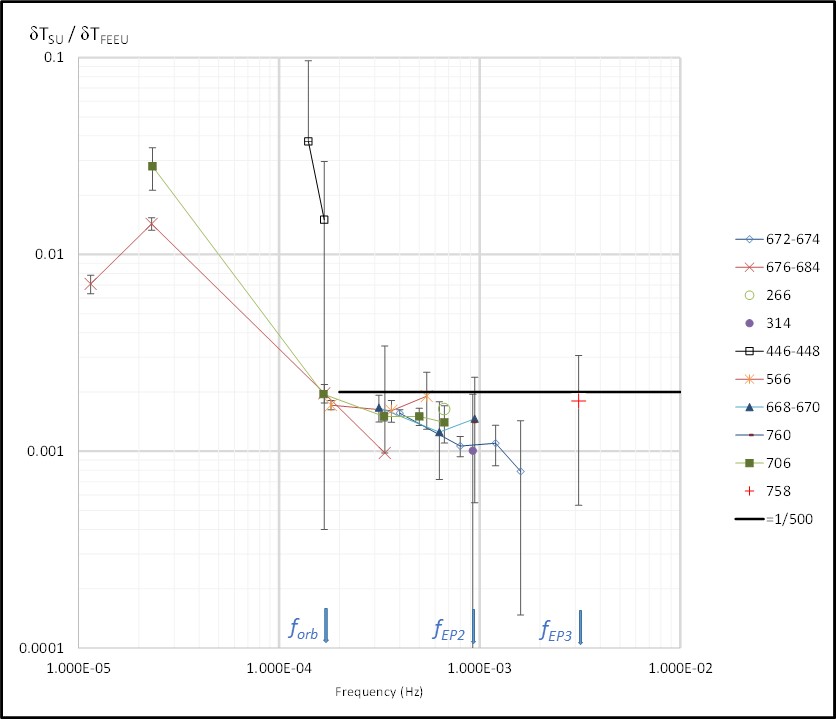}
\caption{Estimation of the thermal filtering between the FEEU stage and the SU stage for different sessions with a temperature stimulus at FEEU stage. The points of the same session are linked by a line.}
\label{fig_temp_exp} 
\end{figure}

For sessions \#668-670 and \#672-674, stimulus signals similar to those used in the thermal sensitivity (\Sref{therm_calib}) have been applied on the platform heaters of the FEEU plate without current compensation. The accelerometer was operating in a non scientific mode of the satellite, mode that does not allow for accurate acceleration measurements. However this mode allows us to acquire the different temperature data to characterise the thermal filtering. A session of 120 orbits (\#668-670) with a stimulus at $3.2\times{}10^{-4}$\,Hz was applied first and then followed by a stimulus at $4\times{}10^{-4}$\,Hz during 150 orbits. The temperatures of the different units were recorded and show significant signals at the stimuli frequencies and their harmonics. The ratio $\delta T_{\rm SU}/\delta T_{\rm FEEU}$ is plotted in \Fref{fig_temp_exp}. 

\Fref{fig_temp_exp} shows that for all sessions, the ratio $\delta T_{\rm SU}/\delta T_{\rm FEEU}$ is lower than $1/500$ for frequencies higher than $f_{\rm orb}$. Session \#758 shows temperature variations at the limit of the noise and the attenuation plotted in this figure is a conservative number and should be much lower (at $3\times{}10^{-3}$\,Hz). This point was nevertheless kept in the graph. A dependency to frequency seems to appear as in \Fref{fig_temp_spicho} which we did note consider in order to be conservative. 

It must be noted that the analysis of EP sessions does not allow for a direct access to the ratio $\delta T_{\rm SU}/\delta T_{\rm FEEU}$ because no temperature signal comes out from the SU temperature probe noise. However a signal in the FEEU temperature measurements is observable and can be used to estimate the ratio $\delta T_{\rm FEEU}/\delta T_{\rm RAD}$ in V2 configuration. In V3 configuration, the FEEU signal is also at the limit of the noise and has been considered as an upper bound. \Tref{tab_ratio} gives a summary of these observations in EP sessions which are in very good agreement with the particular session performed with the tilted satellite (session \#676-684).  It confirms that SUEP and SUREF have a similar thermal behaviour as expected.
The observation of the FEEU temperature in the EP sessions determine the figure to be taken into account in the final budget (see below). The mean temperature variation in the FEEU has been estimated to $375\mu$\textdegree{}C$\pm{}40\mu$\textdegree{}C for SUEP and to $390\mu$\textdegree{}C$\pm{}45\mu$\textdegree{}C for SUREF, at $f_{\rm EP}$, in V2 mode. In V3 mode, the FEEU temperature variations at $f_{\rm EP}$ have been estimated to $70\mu$\textdegree{}C$\pm{}24\mu$\textdegree{}C for SUEP and to $100\mu$\textdegree{}C$\pm{}42\mu$\textdegree{}C for SUREF.

\begin{table}
\caption{{\label{tab_ratio}}Summary of FEEU temperature responses to a radiator temperature variation.}
\begin{tabular}{lccc} 
\textbf{${\Delta T_{\rm RAD}}/{\Delta T_{\rm FEEU}}$} & \textbf{SUREF} & \textbf{SUEP} & extrapolated from \Fref{fig_temp_spicho}       \\ 
\hline
in V2 mode & $34 \pm{}6$ & $37\pm{}4$ & $30\pm{}5$\\
in V3 mode & $89 \pm{}36$ & $115\pm{}43$ & $99\pm{}5$\\
\end{tabular}
\end{table}

\subsection{Synthesis of thermal systematic effects}

The thermal sensitivity is higher than expected prior to the launch in the sensor units and in the FEEU. For the SU, it could be due to its integration in the satellite and the resulting thermal expansion which was underestimated in the budget of the SU considered alone. For the FEEU, the temperature range in flight is lower than the range expected prior to the flight because of operational and safety constraints. The breakdown of a FEEU capacitor linked to SUREF \cite{rodriguescqg4} led to adapt the mission scenario and to have, most of the time, only one SU operating at one given time. This operational constraint which aims to preserve the life time of both FEEU decreases the instrument operational temperature at about 10$^{\circ}$C, lower than the design range of [20$^{\circ}$C-40$^{\circ}$C]. The thermal sensitivity of the capacitive detector was characterised in this design range and an operation out of this range is known to degrade thermal sensitivity.

Fortunately, in-flight experiments for the characterisation of the thermal filtering have shown that:
\begin {itemize}
\item the temperature fluctuation at $f_{\rm EP}$ that generates a systematic error comes from the variation of Earth's albedo entering the FEEU radiator;
\item the temperature fluctuation of the FEEU radiator is transmitted to the FEEU panel and then to the SU panel with an attenuation factor decreasing with frequency;
\item the thermal filtering between the FEEU radiator, the FEEU panel and the SU panel looks like a first-order low-pass filter;
\item SUREF and SUEP behave similarly with respect to temperature filtering;
\item since few data are available for frequencies higher than $10^{-3}$\,Hz, it was decided to use the ratio ${\delta T_{\rm SU}}/{\delta T_{\rm FEEU}}=1/500$, the highest ratio to be the most conservative for frequencies higher than $0.924\times{}10^{-3}$\,Hz.
\end {itemize} 

The total systematic error due to thermal variations $\Gamma^{(d)}_{\rm Tth}$ is given by adding to \Eref{eq_therm} the effect of the temperature sensitivity of calibrated parameters established in \Sref{calib}:
\begin {equation} \label{Tth_sys}
\begin{split}
\Gamma^{(d)}_{\rm Tth}(f_{\rm EP}) &= \left[\lambda_{\rm SU}\delta T_{\rm SU}(f_{\rm EP})+\lambda_{\rm FEEU}\delta T_{\rm FEEU}(f_{\rm EP})\right] \\
& +\left[\frac{\partial a_{d11}}{\partial T_{\rm SU}}\delta T_{\rm SU}(f_{\rm EP}) + \frac{\partial a_{d11}}{\partial T_{\rm FEEU}}\delta T_{\rm FEEU}(f_{\rm EP})\right]\Gamma_x^{(c)}.   
\end{split}
\end {equation}
The sensitivities $\lambda_{\rm SU}$, $\lambda_{\rm FEEU}$ at each frequency test and the ones of $a_{d11}$ are given respectively in \Tref{thresults} and \Tref{tab_ad11}. The mean applied acceleration $\Gamma_x^{(c)}$ is evaluated to the mean bias of the two sensors concerned by the DFACS loop as a good guess of the remaining acceleration in the output measurements. The FEEU temperature variations $\delta T_{\rm FEEU}$ at $f_{\rm EP}$ comes from the probe measurement data in each session. The SU temperature variations, being so low, are estimated through the in-flight model verified in \Sref{flight_therm}: $\delta T_{\rm SU}=\delta T_{\rm FEEU}/500$. Consequently, the disturbing acceleration becomes: 
\begin {equation} \label{th_sys}
\Gamma^{(d)}_{\rm th}(f_{\rm EP}) = \left[\frac{\lambda_{\rm SU}(f_{\rm EP})}{500}+\lambda_{\rm FEEU}(f_{\rm EP}) +\left(\frac{\frac{\partial a_{d11}}{\partial T_{\rm SU}}}{500} + \frac{\partial a_{d11}}{\partial T_{\rm FEEU}}\right)\Gamma_x^{(c)}\right]\delta T_{\rm FEEU}(f_{\rm EP}).   
\end {equation}



\section{Magnetic sensitivity analysis} \label{magn_sys}

	\subsection{Requirements on magnetic effects}

An analytical expression of the magnetic field is hardly possible here, as the resultant effect on a test-mass depends on a very local gradient of field. The magnetic field and its gradient are not uniform in the test-masses area, so that only the integration of local force distribution is relevant. That is why the specification of the magnetic environment has been expressed in terms of disturbing acceleration requirements on the test-masses; these disturbances must comply with the following rules:

\begin{itemize}
\item at DC level, the magnetic effect should induce an acceleration on each test-mass lower than $2.5 \times{}10^{-9}$\,m\,s$^{-2}$ contributing to one tenth of the overall bias;
\item at $f_{\rm EP}$,  the magnetic effect should induce a differential acceleration lower than $2 \times{}10^{-16}$\,m\,s$^{-2}$ in agreement with the error source distribution \cite{rodriguescqg1}, and a common mode acceleration lower than $10^{-12}$\,m\,s$^{-2}$;
\item the acceleration noise around $f_{\rm EP}$ coming from magnetic effect should be lower than $6 \times{}10^{-14}$\,m\,s$^{-2}$\,Hz$^{-1/2}$ to be compatible with the $2 \times{}10^{-16}$\,m\,s$^{-2}$ at $f_{\rm EP}$ after a 20-orbit integration time;
\item at $2f_{\rm EP}$, the magnetic effect should not disturb the in-orbit calibration of the offcentrings (\Sref{Sect_Off}) and should lead to a differential acceleration error lower than $0.4 \times{}10^{-14}$\,m\,s$^{-2}$ in order to contribute to less than $25$\% of the target accuracy on offcentring of 0.1$\mu$m after 20-orbit integration time. The common mode effect should be lower than $2 \times{}10^{-11}$\,m\,s$^{-2}$ at $2f_{\rm EP}$. 
\end{itemize}

In order to verify the requirements, an analysis has been conducted by considering Earth's magnetic field values coming from a data base and a finite element model of the satellite to compute the magnetic effect on the test-masses. This analysis helps also to fix a magnetic moment requirement on each equipment: the noise level and the variation at $f_{\rm EP}$ of the magnetic moments were specified respectively to $4\times{}10^{-2}$\,A\,m$^2$\,Hz$^{-1/2}$ and to $10^{-2}$\,A\,m$^2$.

\subsection{General analysis and equations}

The test-masses made of platinum and titanium alloys undergo the Earth and satellite magnetic field. Although weak but not null the magnetic susceptibility leads to a magnetic force on the conducting test-masses: the induced magnetic moment of each test-masses tends to orientate in the magnetic field environment; the test-mass  travelling in the Earth's magnetic field undergoes a Lorentz force. The Lorentz force can be neglected because the test-mass metallic housing acts as an electrical shield. Indeed, let us consider $\overrightarrow{v_s}$ the velocity of the test-mass in the Earth's reference frame, assumed identical to the velocity of the satellite, and $\overrightarrow{E'}$ and $\overrightarrow{B'_E}$ the electrical and magnetic fields in the Earth's reference frame. The electrical $\overrightarrow{E}$ and magnetic fields $\overrightarrow{B_E}$ in the test-mass reference frame become thus (via a Lorentz transformation): $\overrightarrow{E}=\overrightarrow{E'}- \overrightarrow{v_s} \times{}\overrightarrow{B'_E}$ and $\overrightarrow{B_E}=\overrightarrow{B'_E}$ (since the satellite velocity is much smaller than the speed of light). In the local frame, the metallic vacuum housing provides a Faraday cage to the test-mass and thus $\overrightarrow{E'}=0$ and the test-mass is shielded from a Lorentz force. 
However this shielding and the additional shielding around the payload cannot fully cancel the effect of the magnetic field on the residual test-mass magnetic moment. These forces can create disturbing signals in the EP test data and have to be evaluated and minimised. 
The disruptive force along the $X$ axis is:

\begin{equation}  \label{eq_forceMg}
F_{B,x}=\sum_i \overrightarrow{m_i} \left( \frac{\partial \vec{B_t}}{\partial x} \right)_i+\frac{\chi}{\mu_0}\int_V \vec{B_t} \frac{\partial \vec{B_t}}{\partial x} dV + \left(\int_V \vec{J} \times{}\vec{B_t} dV \right) \overrightarrow{u_x},
\end{equation}
where $B_t$ is the sum of the Earth's and satellite magnetic fields.
Here it is assumed that the diamagnetic test-mass, with susceptibility $\chi$, includes ferromagnetic inclusions and the first term of the equation is the sum over all inclusions. The worst case is the external mass of the SUEP instrument made of titanium alloy TA6V which contains 0.25\% iron. $\overrightarrow{m_i}$ is the permanent magnetic moment of the $i{\rm th}$ ferromagnetic inclusion, and $\left( \frac{\partial \vec{B_t}}{\partial x} \right)_i$ is the magnetic field $x$ derivative at the location of each inclusion. The magnetic field $\vec{B_t}$ includes the Earth ($\overrightarrow{B_E}$) and satellite contribution. $\vec{J}$ is the density of any macroscopic current circulating within the test-masses. The integrals are calculated over the volume $V$ of the test-masses. 

The first two terms of \Eref{eq_forceMg} produce low frequency accelerations (noise and systematic at the EP test frequency) proportional to magnetic field fluctuations at the same frequency and are evaluated with a finite element analysis below. 
The third term was estimated on the basis of the work in Ref. \cite{Ray18, Nagel18}: to calculate the force acting on a metallic element, the study first considers a small conducting sphere placed inside a time-varying magnetic field. Due to the changing magnetic field, an electrical eddy current density is induced throughout its volume, giving rise to a magnetic moment. Combined with the magnetic field gradient, this moment produces a force. This study also shows that for the general case of non-spherical geometries, the method can be applied with good accuracy if an equivalent spherical radius is provided. Even if the purpose of this analysis is dedicated to the electrodynamic sorting of metals and thus to generate a continuous force, the conclusions can be used to estimate the disturbing forces in the case of MICROSCOPE at low frequencies. The analysis concludes that this contribution is smaller by five orders of magnitude than the second term of Eq.\eref{eq_forceMg} and can thus be ignored. The disturbing force along the $X$ axis can be simplified to:  

\begin{equation}  \label{eq_forceMg2}
F_{B,x}=\sum_i \overrightarrow{m_i} \left( \frac{\partial \vec{B_t}}{\partial x} \right)_i+\frac{\chi}{\mu_0}\int_V \vec{B_t} \frac{\partial \vec{B_t}}{\partial x} dV.
\end{equation}

	\subsection{Modelling of the magnetic environment}
In order to estimate magnetic acceleration errors, a good knowledge of the magnetic environment is required. Therefore, the sources of magnetic field that may affect the MICROSCOPE instrument have been analysed. They are either produced by the Earth and depend on the orbit followed or produced by the satellite equipment. Both aspects have been studied and are reported below.

	\subsubsection{Earth's magnetic environment}
The Earth's magnetic field amplitude has been estimated by the IPGP (Paris Institute of Earth Physics) with Oersted satellite data \cite{ipgp} and an extrapolation to MICROSCOPE orbit scenario. This data base allowed us to compute the harmonic levels of the projected Earth's magnetic field onto the axes of the instrument: the values are given in \Tref{Magn_inert} for the inertial mode and in \Tref{Magn_spin} for the satellite rotating mode. Because of the rotation mode about $Y$ axis, the magnetic field along this axis has no component at some frequencies and their corresponding values are not mentioned in \Tref{Magn_spin}, they can be neglected.

\begin{table}[H]
\caption{\label{Magn_inert} Spectral composition evaluation of the Earth's magnetic field in inertial mode projected on the SU reference frame ($f_{\rm EP}=f_{\rm orb}$).}
\begin{tabular}{l|ccc}
Frequencies & $B_{Ex}$(T) & $B_{Ey}$(T)  & $B_{Ez}$(T)        \\ \hline
DC                & $3.1 \times{}10^{-8}$  & $0.3 \times{}10^{-5}$  &  $3.1 \times{}10^{-5}$    \\
$f_{\rm EP}$   & $0.7 \times{}10^{-6}$   & $2.2 \times{}10^{-6}$ & $ 0.5\times{}10^{-6}$ \\
$2f_{\rm EP}$ & $2.2 \times{}10^{-5}$ & $0.1 \times{}10^{-5}$   & $2.2 \times{}10^{-5}$ \\
$3f_{\rm EP}$   &$2.5 \times{}10^{-6}$&  $0.2 \times{}10^{-6}$  & $2.0 \times{}10^{-6}$ \\
\end{tabular}
\end{table}

In the rotating mode, the spin about the $Y$ axis modulates the magnetic field on the orbital plane $(X,Z)$ with the spin rate frequency. 
For the stochastic term, IPGP provided us with data measured by the Oersted satellite whose altitude of 618~km can be considered dimensioning compared to the MICROSCOPE's one. Worst case random noise of the Earth’s magnetic field has been found to be of $104$\,nT\,Hz$^{-1/2}$.

\begin{table}[H]
\caption{{\label{Magn_spin}} Spectral composition evaluation of the Earth's magnetic field in rotating mode projected on the SU reference frame. No value means that the estimation is not relevant at this frequency.}
\begin{tabular}{l|cc}
Frequencies                             & $B_{Ey}$ (T)                        & $B_{Ex}$(T) and $B_{Ez}$(T)   \\ \hline
DC                                    & $0.3 \times{}10^{-5}$   &  $\simeq 0$\\
$f_{\rm EP}-f_{\rm orb}$    &  &$0.8 \times{}10^{-5}$    \\
$f_{\rm orb}$                      &  $2.2 \times{}10^{-6}$  & $0.4 \times{}10^{-6}$\\
$f_{\rm EP}+f_{\rm orb}$   &                                  & $2.5 \times{}10^{-7}$\\
$f_{\rm EP}-2f_{\rm orb}$  &                                   & $0.1 \times{}10^{-6}$\\
$2f_{\rm orb}$                    & $0.1 \times{}10^{-5}$    &  \\
$f_{\rm EP}+2f_{\rm orb}$   &                                   & $1.5 \times{}10^{-7}$ \\
$f_{\rm EP}-3f_{\rm orb}$   &                                  & $2.5 \times{}10^{-5}$ \\
$3f_{\rm orb}$                     & $0.2 \times{}10^{-6}$   & \\
$f_{\rm EP}+3f_{\rm orb}$  &                                  & $0.3 \times{}10^{-7}$\\
\end{tabular}
\end{table}

	\subsubsection{Satellite magnetic environment}
\begin{figure*}
\includegraphics[width=0.5 \textwidth]{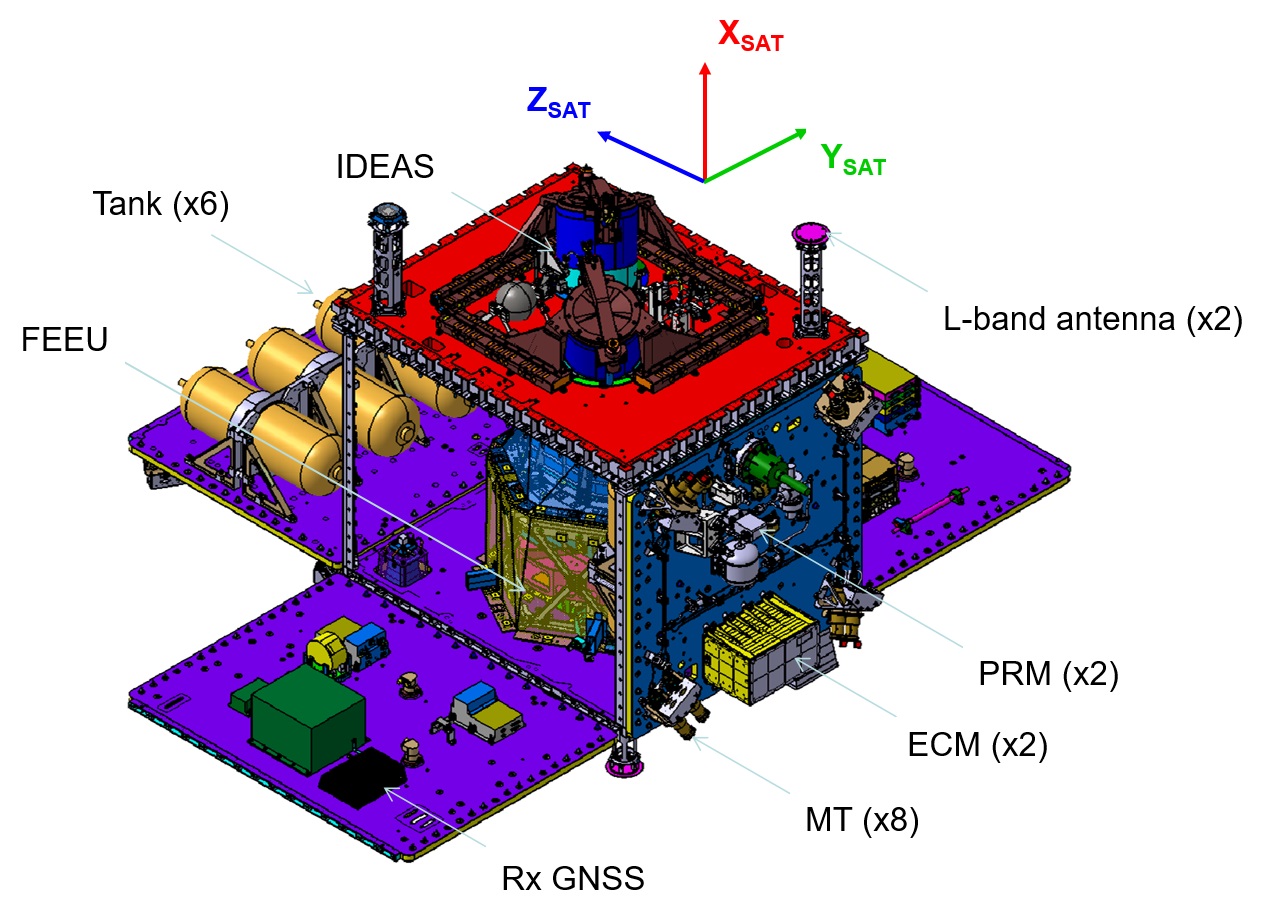}
\includegraphics[width=0.5 \textwidth]{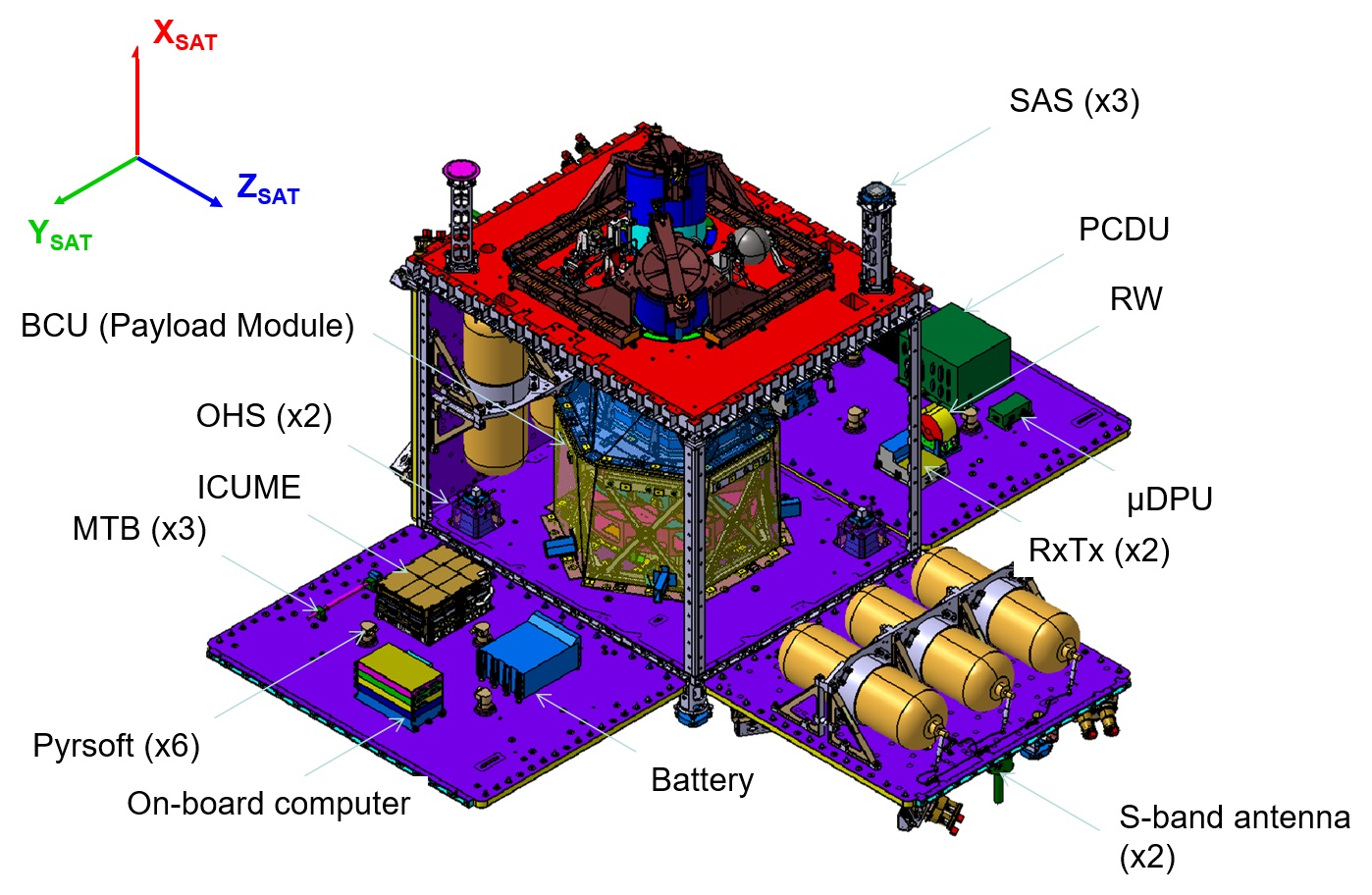}
\caption{Satellite drawing with equipment which may generate a magnetic field. }
\label{fig_sat} 
\end{figure*}

The magnetic equipment of the platform and payload are well identified. Nevertheless, few information on the magnetic characteristics is available for various reasons. First, the information provided by manufacturers is often limited to a maximum value of the magnetic moment. Second, some equipment consists of moving parts whose magnetic moment orientation may thus change (typically the reaction wheel even if they are not used in science mode). Additionally, the magnitude of the magnetic moment may also depend on the electrical current intensity which varies: this is particularly the case for the power conditioning unit. Finally, some materials have magnetic properties such that they generate an induced magnetic field when they undergo a magnetic field. This is particularly the case of the battery.
Thus the magnetic moments of the equipment were estimated either by considering the manufacturer data, or by direct measurements, and completed by simulations. \Tref{Magn_sat} summarises this information for major equipment contributors of the platform schematised in \Fref{fig_sat}. In this table, data are given in the satellite reference frame (\Fref{fig_sat}). 

\begin{table}
\caption{\label{Magn_sat} Magnetic moment distribution in the satellite reference frame. The $\pm{}$ sign means that the direction of the magnetic moment of the equipment is indeterminate, the total is a worst case that takes this into account.}
\begin{tabular}{|l|c|c|c|c|}
\hline
                & Distance &   \multicolumn{3}{c|}{Magnetic moment}       \\
Equipment & to payload &   \multicolumn{3}{c|}{(mA\,m$^2$)}       \\
 & (mm) & $M_{X_{\rm sat}}$ & $M_{Y_{\rm sat}}$ &$M_{Z_{\rm sat}}$         \\
 \hline
 \multicolumn{5}{|c|}{On +$X_{\rm sat}$ panel}\\ 
\hline
$Y_{\rm sat}$ MTB    & 519  & 0 & $\pm{}60$ & $0$\\
$Z_{\rm sat}$ MTB   & 580  & 0 & $0$ & $\pm{}60$ \\
IDEAS    &572  & $\pm{}$58 & $\pm{}$58 & $\pm{}$58\\
\hline
 \multicolumn{5}{|c|}{On -$X_{\rm sat}$ panel}\\ 
\hline
Star sensor head 1   & 554  & $\pm{}$5 & -55 & $\pm{}$5\\
Star sensor head 2  & 683  & $\pm{}$5 & $ \pm{}$5 & -55 \\
\hline
 \multicolumn{5}{|c|}{On +$Y_{\rm sat}$ panel}\\ 
\hline
GS +$Y_{\rm sat}$ panel  & 1047  & -10 & 0 & 0\\
$X_{\rm sat}$ MTB  & 562  & $\pm{}$60 & 0 & 0 \\
OBC  & 431  &3 & -3 & -5 \\
Battery  & 412  &139 & 113 & 122 \\
\hline
 \multicolumn{5}{|c|}{On -$Y_{\rm sat}$ panel}\\ 
\hline
GS -$Y_{\rm sat}$ panel  & 1047  & -10 & 0 & 0\\
$X_{s\rm at}$ reaction wheel  & 409  & 4 & $\pm{}9$ & $\pm{}$9 \\
Star sensor electronics  & 509  & 5 & 5 & -5 \\
PCDU  & 414  & -10 & -5 & -100 \\
\hline
 \multicolumn{5}{|c|}{On +$Z_{\rm sat}$ panel}\\ 
\hline
PRM   & 567  &  101 & -420 & 953 \\
ECM  & 618  & $\pm{}$17 & $\pm{}$17 & $\pm{}$17 \\
\hline
 \multicolumn{5}{|c|}{On -$Z_{\rm sat}$ panel}\\ 
\hline
PRM   & 590  &  101 & 342 & -862 \\
ECM  & 640  & $\pm{}$17 & $\pm{}$17 & $\pm{}$17 \\
\hline
 \multicolumn{5}{|c|}{Payload}\\ 
\hline
ICUME on -$X_{\rm sat}$   & 427  & $\pm{}$6 & $\pm{}$6 & $\pm{}$6  \\
Along +$Z_{\rm sat}$ FEEU   & 288  & $\pm{}$6 & $\pm{}$6 & $\pm{}$6  \\
Along -$Z_{\rm sat}$ FEEU  & 295  & $\pm{}$6 & $\pm{}$6 & $\pm{}$6  \\
\hline
 \multicolumn{2}{|c|}{Total worst case} & 502 & -443 & -380 \\
\hline
\end{tabular}
\end{table}

	\subsubsection{Test-mass magnetic measurement}
The permanent magnetic moment of the test-mass appears in the first term of \Eref{eq_forceMg2}. The measurement was made by PTB (Physikalisch Technische Bundesanstalt, the National Metrology Institute of Germany) and the magnetic moment was estimated to be less than 30\,nA\,m$^2$ (noise limit of the measurement).

\subsection{Finite element analysis}
\begin{figure*}
\includegraphics[width=0.5 \textwidth]{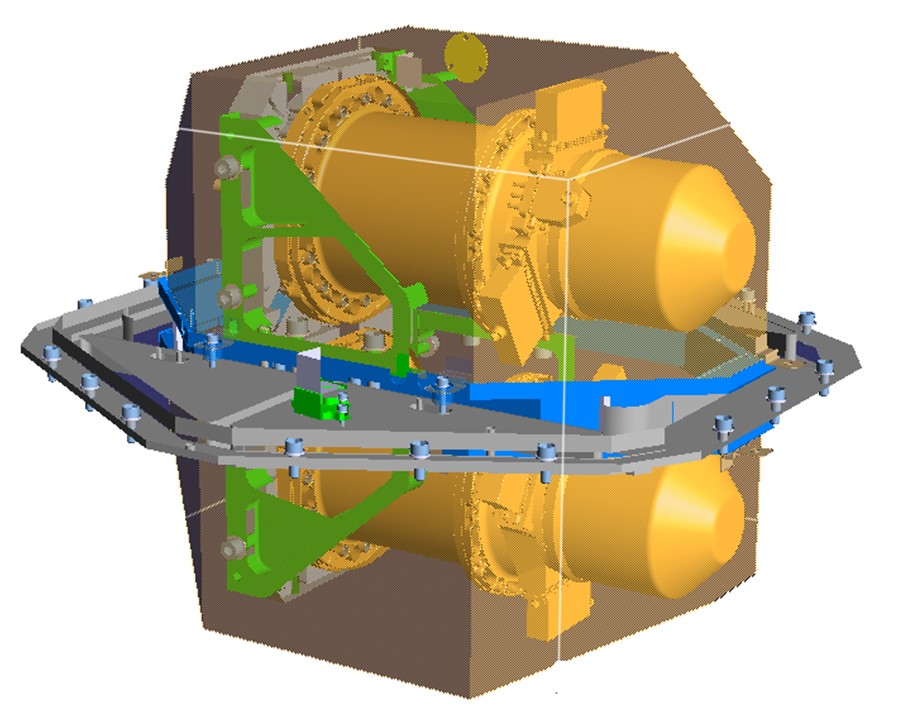}
\includegraphics[width=0.5 \textwidth]{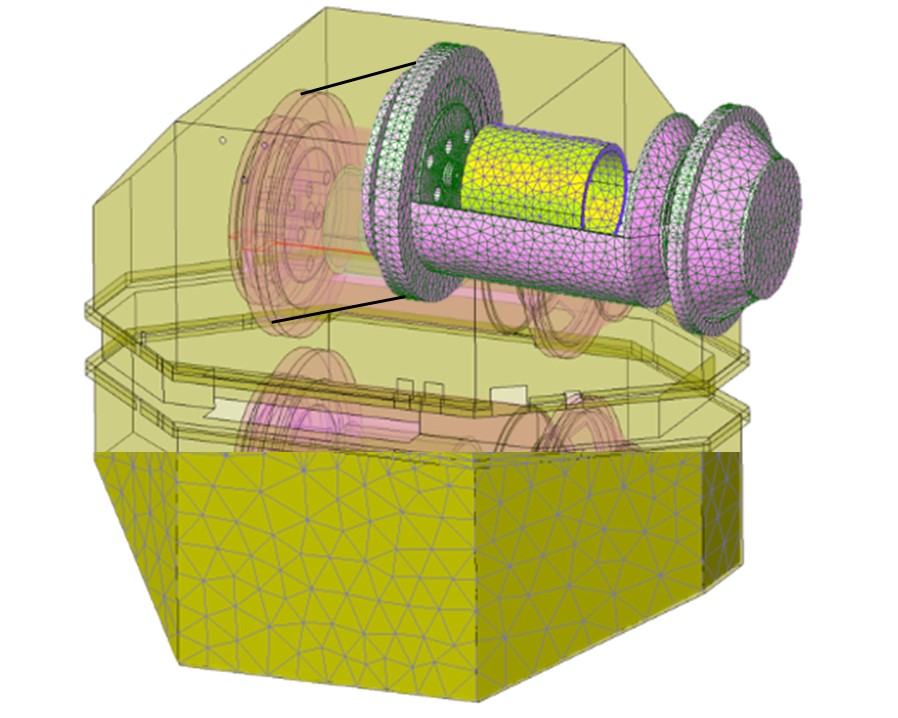}
\caption{Left: the instrument and its external shielding which surrounded only the sensor units. Right: overview of the instrument meshing and of the shielding meshing, the SU has a much finer meshing to better estimate local gradients.}
\label{fig_shield} 
\end{figure*}

The model takes into account both shielding: one around the mechanical core provided by the vacuum housing made of Invar\textregistered{} and an additional one around the payload case (\Fref{fig_shield}) provided by sheets of 8\,mm made of Supranhyster 50\textregistered{} with high relative permeability ($\mu_r>$10000).

The Invar material around the SU and the additional shielding in Supranhyster have a major impact in the magnetic field environment that makes finite element model necessary to compute the magnetic field and its gradients. The method used was first to determine by numerical simulation the field and field gradient mapping in the test-masses for two sources of magnetic disturbance:
\begin {itemize}
\item a uniform field in the three directions to simulate the Earth’s magnetic field;
\item a near magnetic dipole (30 cm) to simulate the effect of a satellite equipment and oriented in three directions (considered as a dimensioning case).
\end {itemize}

In the end, therefore, six maps are available at the level of the test-masses. 
The residual magnetic field at the test-masses is calculated on three 2D grids (per mass) with an angular pitch of 20 degrees or 18 points and 16 points along the length for the external mass and 11 for the internal mass.

\subsection {Synthesis of magnetic systematic effects.}
\Eref{eq_forceMg2} has been integrated numerically on each mass. It was thus possible to obtain an estimate of the magnetic disturbance and its spectral composition in inertial and spinning mode.

The upper part of the \Tref{mag_1} gives the contributions of magnetic effect to the error budget which corresponds to the first term of \Eref{eq_forceMg2}: $\frac {1}{m} \sum_i \overrightarrow{m_i} \left( \frac{\partial \vec{B_t}}{\partial x} \right)_i$ due to the magnetic moments distributed on the satellite with $m$ the mass of the test-mass.
The middle part of the \Tref{mag_1} gives the contributions of magnetic effect to the error budget which correspond to the second term of \Eref{eq_forceMg2}: $\frac {1}{m} \frac{\chi}{\mu_0}\int_V \vec{B_t} \frac{\partial \vec{B_t}}{\partial x} dV$due to the test-mass magnetic susceptibility. 

\begin{table} [h]
\caption{\label{mag_1} Summary of the acceleration disturbances on each SU due magnetic sensitivity and total of systematic error at $f_{\rm EP}$.}
\begin{tabular}{l|cccc}
\hline
 \multicolumn{5}{c} {Magnetic disturbances due to distributed magnetic moments}\\
\hline
 & Bias & Noise & \multicolumn{2}{c} {Variations}      \\ 
 Sat. mode & at DC  & around $f_{\rm EP}$ & at $f_{\rm EP}$ & at $2f_{\rm EP}$        \\ 
  & (m\,s$^{-2}$) & (m\,s$^{-2}$\,Hz$^{-1/2}$) & \multicolumn{2}{c} {(m\,s$^{-2}$)}        \\ 
\hline
\textbf{On SUREF} &&&&\\
Inertial mode   & $2.7 \times{}10^{-16}$  & $1.4 \times{}10^{-17}$ & $2.9 \times{}10^{-19}$ & $9.6 \times{}10^{-20}$\\
Spin mode    & $2.7 \times{}10^{-16}$  & $1.4 \times{}10^{-17}$ & $2.8 \times{}10^{-19}$ &0\\
\hline
\textbf{On SUEP}&&&&\\
Inertial mode   & $7.2 \times{}10^{-15}$  & $4.1 \times{}10^{-16}$ & $7.4 \times{}10^{-18}$ & $3.5 \times{}10^{-19}$\\
Spin mode    & $7.2 \times{}10^{-15}$  & $4.1 \times{}10^{-16}$ & $7.2 \times{}10^{-18}$ &0\\
\hline
 \multicolumn{5}{c} {Magnetic disturbances due to test-mass susceptibility}\\
\hline
 Sat. mode & at DC  & around $f_{\rm EP}$ & at $f_{\rm EP}$ & at $2f_{\rm EP}$        \\ 
  & (m\,s$^{-2}$) & (m\,s$^{-2}$\,Hz$^{-1/2}$) & \multicolumn{2}{c} {(m\,s$^{-2}$)}        \\ 
\hline
 \textbf{On SUREF}&&&&\\
Inertial mode   & $1.4 \times{}10^{-16}$  & $1.7 \times{}10^{-16}$ & $1.5 \times{}10^{-16}$ & $1.4 \times{}10^{-16}$\\
Spin mode    & $1.5 \times{}10^{-16}$  & $1.7 \times{}10^{-16}$ & $4.6 \times{}10^{-18}$ &$5.4 \times{}10^{-19}$\\
\hline
 \textbf{On SUEP}&&&&\\
Inertial mode   & $2.4 \times{}10^{-16}$  & $2.1 \times{}10^{-16}$ & $1.9 \times{}10^{-16}$ & $2.3 \times{}10^{-16}$\\
Spin mode    & $2.5 \times{}10^{-16}$  & $2.1 \times{}10^{-16}$ & $7.9 \times{}10^{-18}$ &$8.1\times{}10^{-19}$\\
\hline
 \multicolumn{5}{c} {Total systematic error in spin mode at $f_{\rm EP}$ (m\,s$^{-2}$)}\\
\hline
 \textbf{On SUREF}& & & $0.6 \times{}10^{-18}$ &\\
 \textbf{On SUEP}& & & $14.4 \times{}10^{-18}$ & \\
\end{tabular}
\end{table}

Since this is a worst case, there is no difference between the common and the differential modes. All errors are algebraically added without taking into account the phase of the signals, especially for the systematic error at $f_{\rm EP}$. Even in this context, the error remains lower than the specification in inertial mode and is two orders of magnitude lower in spin mode.


\section{Local gravity effect} \label{grav_sys}

\begin{figure*}
\includegraphics[width=0.5 \textwidth]{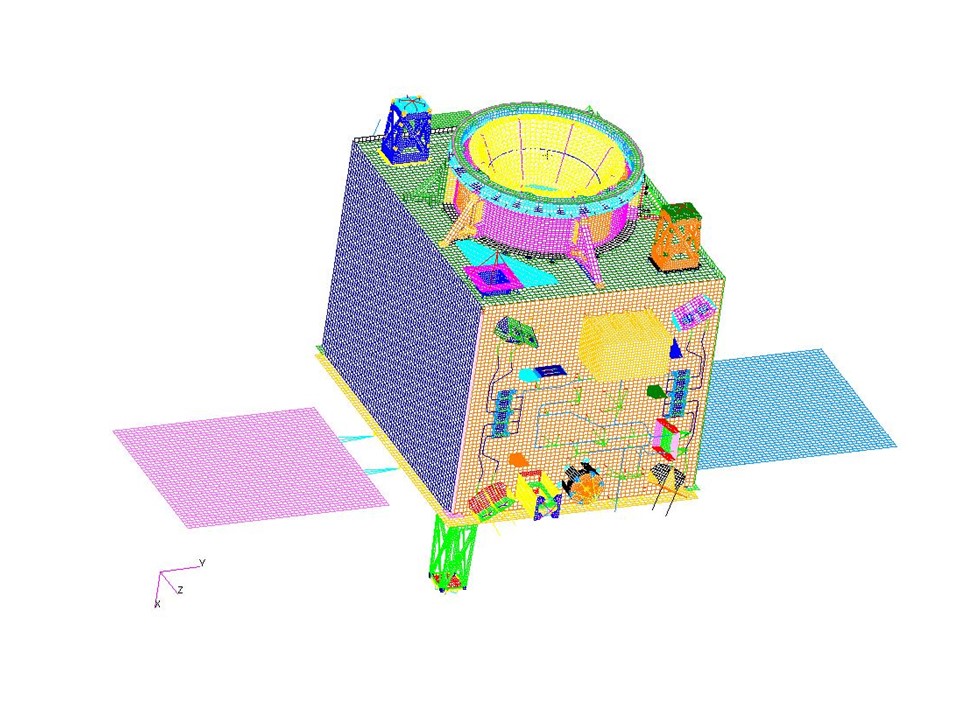}
\includegraphics[width=0.5 \textwidth]{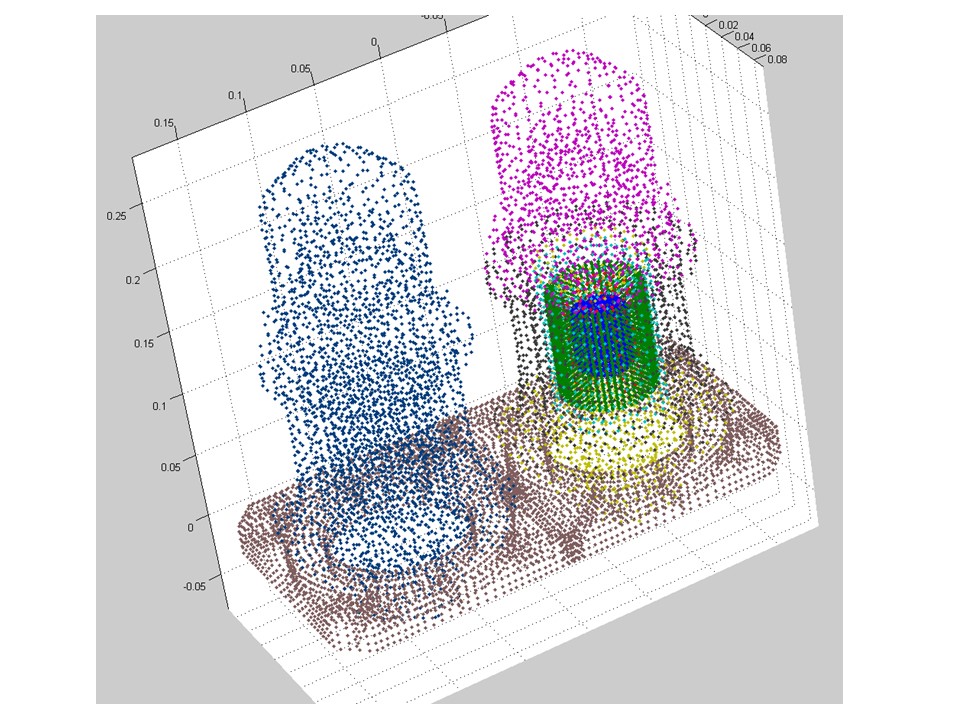}
\caption{Meshing of the satellite (left) and of the sensor unit (right) for the calculation of the local gravity impact on the acceleration measurement. }
\label{fig_gravm} 
\end{figure*}

The local gravity field has been evaluated by two fine finite element models (\Fref{fig_gravm}): one modelling each piece of equipment of the satellite except the instrument, and a second one modelling the two sensor units surrounding the test-masses. We have considered the gravity variations due to thermal expansion of the satellite or the instrument. These gravity variations affect differently the two test masses of a same SU in two different ways: either due to the difference of positions of their centres (gravity gradient) or due to their difference of shape. The difference of shape has an effect on the moment of inertia about all axes. Ideally identical moments of inertia about all axes make the test-mass behaving as a sphere for the gravity field up to second order \cite{rodriguescqg1}. Defects to the ideal shape break the symmetry of the spherical inertia and lead to a difference of gravity field undergone by each test-mass even if they are perfectly centred (from a geometrical point of view).

In the case of the satellite model, the gravity field of each element is calculated at each SU test-mass centre position. Several scenarios have been studied to cover different periods of the year and to take into account the two modes of the satellite in science mode: inertial pointing and rotating. These different cases help to identify the worst case for thermoelastic displacement of masses on board the satellite by considering the variations of temperature in the different scenarios. The in-flight data analysis shows a very good agreement with the satellite thermal model and appears to be much better in rotating mode (\Sref{therm_sys}). \Tref{grav_1} gives the results of the finite element analysis of the satellite for the worst case (inertial pointing) of the year. The table includes also the effect of the test-mass shape with moments of inertia not identical about all axes by 0.1\% (worst case value of the four test-masses \cite{touboul19}).

In the case of the instrument model, several scenarios have also been considered in order to identify the worst case of moving parts. Due to the very stable temperature in all satellite configurations as detailed in \Sref{flight_therm}, the worst case is the inertial pointing mode with temperature variation lower than 100$\mu$K which is taken into account to estimate the impact. The test-masses of each SU are considered at the centre of the finite element model and all the other parts are considered to be expanding with respect to this centre as temperature fluctuates. As for the satellite, the instrument thermoelastic model gives all the information to calculate the induced gravity field variations and its gradient. The total induced acceleration is summarised in \Tref{grav_1}. The two error posts (instrument and satellite gravity impacts) are taken into account in the final summary table of systematic errors in \Sref{sum_sys}.

\begin{table}
\caption{\label{grav_1} Results of the finite element study, computation of the gravity variation worst case on board MICROSCOPE satellite at $f_{\rm EP}$ on each SU when considering the gravity source in the satellite and in the SU. The gravity gradient variation contributions are included.}
\begin{tabular}{l|cc}
Gravity source & SUREF & SUEP    \\ 
  variations at $f_{\rm EP}$  & in m\,s$^{-2}$ & in m\,s$^{-2}$     \\ 
\hline

\bf{Satellite thermoelastic} & $1.3 \times{}10^{-16}$  & $0.2 \times{}10^{-16}$ \\
\hline
\bf{Instrument thermoelastic}  & $2.0 \times{}10^{-17}$  & $1.1 \times{}10^{-17}$ \\
\hline
\end{tabular}
\end{table}

\section{Total of systematic errors} \label{sum_sys}

The contribution of each source to the systematic error was evaluated in the previous sections. In this section, we gather those evaluations with other instrument characterisation (bias, angular to linear coupling coefficients \cite{liorzoucqg2, chhuncqg5}). The DFACS \cite{robertcqg3} performance is evaluated a posteriori and inserted in the science data \cite{rodriguescqg4}: it contains the angular control and drag compensation systematic errors inferred from the accelerometer and the star-tracker measurements outputs. For each session the systematic error contributors are evaluated in \Tref{tab_total1} and \Tref{tab_total2}. These contributors have been numbered to facilitate the reading of this section. The synthesis of the systematic errors  over all science sessions is presented in \Tref{tab_budgetF}.
For each error source $k$ (the k$th$ column in \Tref{tab_total1} and \Tref{tab_total2} or line in \Tref{tab_budgetF}), the acceleration $\Gamma^{(d)}_{k,l}$ resulting from the systematic effect in the $l$th session is evaluated. The error groups are the same as the one made in the mission analysis in Ref. \cite{rodriguescqg1}. 

The ``Earth's gravity gradient''contributor in \Tref{tab_budgetF} corresponds to the error correction of the effects listed in the first five lines in \Tref{tab_param} for each EP session $l$. The calibration correction errors have been estimated for each session $l$ and writes:
\begin{equation}
\Gamma_{1,l}^{(d)}= |T_{xx}| \sigma_{\tilde{a}_{c11}\Delta_x}+|T_{xz}| \sigma_{\tilde{a}_{c11}\Delta_z}+|T_{zz}|\tilde{a}_{c13} \sigma_{\tilde{a}_{c11}\Delta_z} +|T_{xy}| \sigma_{\tilde{a}_{c11}\Delta_y}+|T_{yy}| \tilde{a}_{c12} \sigma_{\tilde{a}_{c11}\Delta_y} ,
\end{equation}
where only the components at $f_{\rm EP}$ of $T_{ij}$ are considered, $\sigma_{\tilde{a}_{c11}\Delta_j}$ is the error of the offcentring estimation $\tilde{\Delta_j}$. The model error of the Earth's gravity tensor is negligible here.

The ``instrument and satellite gravity'' contributor, respectively $\Gamma_{2,l}^{(d)}$ and $\Gamma_{3,l}^{(d)}$, have been computed in \Tref{grav_1}. For this contributor, our analysis provides the same upper bound for all sessions.

The ``angular motion'' contributor is determined by the components at $f_{\rm EP}$ of the matrix $[{\rm In}]$ in \Eref{eq_xacc} for each EP session $l$. These terms are also corrected in the data process with the calibrated offcentring. The a posteriori estimation at $f_{\rm EP}$ of the matrix $[{\rm In}]$ is noted $[\tilde{{\rm In}}]$ and leads to a systematic error:  
\begin{equation}
\Gamma_{4,l}^{(d)}=([\tilde{{\rm In}}][\tilde{A_c}]\vv{\tilde{\Delta}})_l+([{\rm In}]\vv{\sigma_\Delta})_l,
\end{equation} 
where $[\tilde{A_c}]$ is the matrix with the $\tilde{a}_{cij}$ terms \cite{rodriguescqg1}. 
It must be noted that the components of $[{\rm In}]$ have been considered without their sign, leading to a conservative estimation for each sessions.

The ``instrument parameter variations'' contributor adresses the time variation of the satellite pointing and of the common mode test-mass alignment, and the angular to linear coupling impact. That does not involve temperature variations of the calibrated parameters included in the next item. The satellite alignment $\alpha_s$ is ensured by the DFACS with a stability better than $1\mu$rad at $f_{\rm EP}$ in each session. The common-mode test-mass alignment $\alpha_{\rm tm}$ is ensured by the accelerometer servoloop wih a stability better than $3\times{}10^{-11}$rad at $f_{\rm EP}$. These two angles project on the $X$ axis the instrument acceleration measurement bias components ($B_{0y}^{(d)}$, $B_{0z}^{(d)}$). The accelerometer bias has been characterised in Ref. \cite{chhuncqg5}. The last part of the ``instrument parameter variation'' contributor comes from the couplings ($c'_{d1m}$) established in Ref. \cite{chhuncqg5} which project the angular acceleration. The levels of the angular acceleration are controlled by the DFACS and given for each session $i$. The ``instrument parameter variation'' is thus computed as: 
\begin {equation}
\begin {split}
\Gamma_{5,l}^{(d)} &= \left(\alpha_s(f_{\rm EP})+\alpha_{\rm tm}(f_{\rm EP})\right)(|B_{0y}^{(d)}|+|B_{0z}^{(d)}|)_l \\
&+2 \left(c'_{d11} |\dt{\Omega}_x(f_{\rm EP})|_l + c'_{d12} |\dt{\Omega}_y(f_{\rm EP})|_l + c'_{d13} |\dt{\Omega}_z(f_{\rm EP})|_l  \right).
\end {split}
\end {equation}

The ``temperature variation'' contributor $\Gamma_{6,l}^{(d)}$, detailed in \Sref{therm_sys}, is given by \Eref{th_sys} computed for each session $l$.

The ``drag-free residuals'' contributor is evaluated with the calibrated instrument parameters $(a_{d1j}/a_{cjj})$ associated to each session $l$.  The systematic error is hence given by:
\begin{equation}
\Gamma_{7,l}^{(d)}=2\left( \frac{a_{d11}}{a_{c11}} n_x^{(c)} + \frac{a_{d12}}{a_{c22}} n_y^{(c)} + \frac{a_{d13}}{a_{c33}} n_z^{(c)}\right)_l.
\end{equation}

The ``magnetic  sensitivity'' contributor $\Gamma_{8,l}^{(d)}$ has been computed in \Tref{mag_1} and has been set to the same value in each session.

The ``non linearity'' contributor has been computed with \Eref{eq_q}. In this expression, the common mode quadratic term $K_{2c,xx}$ has been evaluated in Ref. \cite{chhuncqg5}. But as shown in \Sref{calib}, $K_{2d,xx}$ can vary a lot from session to session and thus could also lead to a variability of $K_{2c,xx}$. As this term is not calibrated regularly on the contrary to $K_{2d,xx}$, the a priori requirement of $14000$\,m$^{-1}$\,s$^{2}$ has been considered as a conservative approximation. This hypothesis is consistent with the estimated $K_{2c,xx}$ in Ref. \cite{chhuncqg5} increased by the standard deviation of $K_{2d,xx}$ estimations. The common and differential accelerations have been approximated by the instrument biases and the DFACS residual accelerations. The resulting estimation is hence given by:
\begin{equation}
\Gamma_{9,l}^{(d)}=2\left[14000 ((n_x^{(c)})_l |B_{0x}^{(d)}|)_l+2(|K_{2d,xx}|_l (n_x^{(c)})_l |B_{0x}^{(d)}|)_l\right],
\end{equation}
where $B_{0x}^{(d)}$ is the differential accelerometric measurement bias.
In this estimation, only absolute values of the components are considered and the $(O)_l$ terms stand for $(O)$ values associated to the session $l$. 

In \Tref{tab_total1} and \Tref{tab_total2}, the contribution of each source $k$ to the systematic error on the E\"otv\"os parameter is obtained by a weighted mean, of the errors related to each session $l$ \cite{metriscqg9}. Sessions contribute with different weights to the estimation of the E\"otv\"os parameter. In Ref. \cite{metriscqg9} we show that combining individual sessions with weights based on their estimation accuracy gives a result very close to the result obtained by directly processing in one step all sessions. Therefore, we consider that each session $l$ may also contribute to the overall systematic error with this weighting:
\begin {equation} \label{eq_tot}
\Gamma^{(d)}_{k}=\frac{1}{\sum_l \frac{1}{\sigma^2_l}} \sum_l \frac{1}{\sigma^2_l} \Gamma^{(d)}_{k,l},
 \end{equation}
where $\sigma_l$ is the $1\sigma$ uncertainty of the E\"otv\"os parameter estimation in the $l$th session. 

\begin{table} 
\caption{\label{tab_total1} SUREF: systematic differential acceleration error calculated per session in rows and per group of errors in columns (detailed in \Sref{sum_sys}). The last column represents the $1\sigma$ uncertainty of the E\"otv\"os parameter estimation on each session, some sessions have been split in two segments  \cite{metriscqg9}.}
    \centering
    \begin{tabular}{|l|l|l|l|l|l|l|l|l|l|r|}
    \hline
          & $\Gamma^{(d)}_{1,l}$ & $\Gamma^{(d)}_{2,l}$ &$\Gamma^{(d)}_{3,l}$ & $\Gamma^{(d)}_{4,l}$ & $\Gamma^{(d)}_{5,l}$ & $\Gamma^{(d)}_{6,l}$ & $\Gamma^{(d)}_{7,l}$ & $\Gamma^{(d)}_{8,l}$ & $\Gamma^{(d)}_{9,l}$ &  $\sigma_{\rm EP}$ \\ \hline
        Session &  \multicolumn{9}{c|}{$\times{}10^{-15}$\,m\,s$^{-2}$} &    $\times{}10^{-15}$  \\ \hline
        120-1 & 0.00 & 0.02 & 0.13 & 0.77 & 1.42 & 43.01 & 0.00 & 0.00 & 0.83 &   16.70 \\ \hline
        120-2 & 0.00 & 0.02 & 0.13 & 0.76 & 1.42 & 43.01 & 0.00 & 0.00 & 0.83 &   8.50 \\ \hline
        174 & 0.00 & 0.02 & 0.13 & 0.23 & 0.13 & 31.59 & 0.01 & 0.00 & 2.80 &   4.90 \\ \hline
        176 & 0.00 & 0.02 & 0.13 & 0.11 & 0.08 & 37.34 & 0.01 & 0.00 & 2.77 &   5.50 \\ \hline
        294 & 0.00 & 0.02 & 0.13 & 0.12 & 0.09 & 15.67 & 0.01 & 0.00 & 3.08 &   2.60 \\ \hline
        376-1 & 0.00 & 0.02 & 0.13 & 0.08 & 0.06 & 29.26 & 0.01 & 0.00 & 3.08 &   7.20 \\ \hline
        376-2 & 0.00 & 0.02 & 0.13 & 0.08 & 0.06 & 29.26 & 0.01 & 0.00 & 3.08 &   6.10 \\ \hline
        380-1 & 0.00 & 0.02 & 0.13 & 0.02 & 0.05 & 5.59 & 0.01 & 0.00 & 3.39 &   3.00 \\ \hline
        380-2 & 0.00 & 0.02 & 0.13 & 0.02 & 0.05 & 5.59 & 0.01 & 0.00 & 3.39 & 3.10 \\ \hline
        452 & 0.00 & 0.02 & 0.13 & 0.03 & 0.07 & 16.26 & 0.02 & 0.00 & 3.71 &   4.00 \\ \hline
        454 & 0.00 & 0.02 & 0.13 & 0.02 & 0.07 & 21.18 & 0.02 & 0.00 & 3.44 &   2.90 \\ \hline
        778-1 & 0.00 & 0.02 & 0.13 & 0.25 & 0.09 & 23.93 & 0.01 & 0.00 & 2.61 &  4.50 \\ \hline
        778-1 & 0.00 & 0.02 & 0.13 & 0.25 & 0.09 & 23.93 & 0.01 & 0.00 & 2.61 &  6.00 \\ \hline
        {\bf Total} & {\bf 0.0} & {\bf 0.0} & {\bf 0.1} & {\bf 0.1} & {\bf 0.1} & {\bf 17.9} & {\bf 0.0} & {\bf 0.0} & {\bf 3.1} & \\ \hline

    \end{tabular}
\end{table}

\begin{table} 
\caption{\label{tab_total2} SUEP: systematic differential acceleration error calculated per session in rows and per group of errors in columns (detailed in \Sref{sum_sys}). The last column represents the $1\sigma$ uncertainty of the E\"otv\"os parameter estimation on each session, some sessions have been split in two segments \cite{metriscqg9}.}
    \centering
    \begin{tabular}{|l|l|l|l|l|l|l|l|l|l|r|}
    \hline
          &  $\Gamma^{(d)}_{1,l}$ & $\Gamma^{(d)}_{2,l}$ &$\Gamma^{(d)}_{3,l}$ & $\Gamma^{(d)}_{4,l}$ & $\Gamma^{(d)}_{5,l}$ & $\Gamma^{(d)}_{6,l}$ & $\Gamma^{(d)}_{7,l}$ & $\Gamma^{(d)}_{8,l}$ & $\Gamma^{(d)}_{9,l}$   &  $\sigma_{\rm EP}$ \\ \hline
        Session &  \multicolumn{9}{c|}{$\times{}10^{-15}$\,m\,s$^{-2}$} &   $\times{}10^{-15}$   \\ \hline
        210 & 0.00 & 0.01 & 0.08 & 0.10 & 0.20 & 13.23 & 0.01 & 0.01 & 6.16 &   14.50 \\ \hline
        212 & 0.00 & 0.01 & 0.08 & 0.08 & 0.11 & 4.41 & 0.01 & 0.01 & 6.82 &   13.90 \\ \hline
        218 & 0.00 & 0.01 & 0.08 & 0.07 & 0.14 & 6.60 & 0.01 & 0.01 & 5.22 &   8.70 \\ \hline
        234 & 0.00 & 0.01 & 0.08 & 0.07 & 0.11 & 6.19 & 0.01 & 0.01 & 5.33 &   9.30 \\ \hline
        236 & 0.00 & 0.01 & 0.08 & 0.01 & 0.07 & 8.21 & 0.03 & 0.01 & 5.47 & 7.40 \\ \hline
        238 & 0.00 & 0.01 & 0.08 & 0.04 & 0.12 & 7.72 & 0.03 & 0.01 & 5.32 &  7.80 \\ \hline
        252 & 0.00 & 0.01 & 0.08 & 0.07 & 0.20 & 6.59 & 0.01 & 0.01 & 5.72 &   8.70 \\ \hline
        254 & 0.00 & 0.01 & 0.08 & 0.20 & 0.31 & 9.91 & 0.01 & 0.01 & 6.11 &   8.40 \\ \hline
        256 & 0.00 & 0.01 & 0.08 & 0.58 & 0.71 & 6.59 & 0.01 & 0.01 & 5.75 &  8.60 \\ \hline
        326-1 & 0.00 & 0.01 & 0.08 & 0.02 & 0.10 & 6.59 & 0.07 & 0.01 & 10.48 &  9.60 \\ \hline
        326-2 & 0.00 & 0.01 & 0.08 & 0.02 & 0.10 & 6.59 & 0.07 & 0.01 & 10.48 &  15.40 \\ \hline
        358 & 0.00 & 0.01 & 0.08 & 0.03 & 0.08 & 6.59 & 0.00 & 0.01 & 5.84 &   11.90 \\ \hline
        402 & 0.00 & 0.01 & 0.08 & 0.09 & 0.15 & 57.25 & 0.01 & 0.01 & 5.73 &   35.10 \\ \hline
        404 & 0.00 & 0.01 & 0.08 & 0.10 & 0.22 & 5.54 & 0.01 & 0.01 & 5.51 &   7.90 \\ \hline
        406 & 0.00 & 0.01 & 0.08 & 0.06 & 0.15 & 24.25 & 0.01 & 0.01 & 6.09 &  18.60 \\ \hline
        438 & 0.00 & 0.01 & 0.08 & 0.10 & 0.38 & 43.02 & 0.01 & 0.01 & 4.87 & 29.60 \\ \hline
        442 & 0.00 & 0.01 & 0.08 & 0.09 & 0.11 & 57.26 & 0.01 & 0.01 & 4.56 &  19.00 \\ \hline
        748 & 0.00 & 0.01 & 0.08 & 0.03 & 0.13 & 57.26 & 0.01 & 0.01 & 4.76 &  59.60 \\ \hline
        750 & 0.00 & 0.01 & 0.08 & 0.13 & 0.34 & 57.26 & 0.01 & 0.01 & 5.22 &  42.40 \\ \hline
        {\bf Total} & {\bf 0.0} & {\bf 0.0} & {\bf 0.1} & {\bf 0.1} & {\bf 0.2} & {\bf 9.3} & {\bf 0.0} & {\bf 0.0} & {\bf 6.0} & \\ \hline

    \end{tabular}
\end{table}

The synthesis of the systematic errors is given in \Tref{tab_budgetF} by adding quadratically all errors \cite{jcgm}: $\sqrt{\sum_k (\Gamma^{(d)}_{k})^2}$ as error sources can be considered independent of each other.
\Tref{tab_budgetF} shows that the drag-free operation of the satellite and the calibration of the instrument have better performances than expected in the design. The environment (thermal, drag-free, magnetism, self-gravity) is better than the specification because of the particular care taken to design the mission. However due to the instrument sensitivity, the main contributor is the temperature variation impact on the differential acceleration measurement. The non-linearity is the second main contributor because of a larger bias of the accelerometer.  
The resulting systematic error on the estimation of E\"otv\"os parameter finally achieved is evaluated to be less than $1.5\times{}10^{-15}$, which is a factor six improvement with respect to the previously published results in \cite{touboul19}.

\begin{table}
\caption{\label{tab_budgetF} Budget of systematic error analysis compared to specification analysis \cite{rodriguescqg1}.}
\begin{indented}
\item[]\begin{tabular}{@{}llccc}
     \br
     &\textbf{Systematic error} &  \textbf{SUEP}   &  \textbf{SUREF} & \textbf{Specification} \\
    & \textbf{sources}              & m\,s$^{-2}$   & m\,s$^{-2}$ & m\,s$^{-2}$\\  
                        
     \mr
   $\Gamma^{(d)}_{1}$ &Earth gravity gradients  & $0.0\times{}10^{-15}$ & $0.0\times{}10^{-15}$ & $0.0\times{}10^{-15}$\\
     \mr
     $\Gamma^{(d)}_{2}$ &Instrument gravity    & $0.0\times{}10^{-15}$ & $0.0\times{}10^{-15}$ & $0.2\times{}10^{-15}$\\
     \mr
     $\Gamma^{(d)}_{3}$ &Satellite gravity gradients   & $0.1\times{}10^{-15}$ & $0.1\times{}10^{-15}$ & $0.3\times{}10^{-15}$\\
     \mr
     $\Gamma^{(d)}_{4}$ &Angular motions  & $0.1\times{}10^{-15}$ & $0.1\times{}10^{-15}$ & $1.1\times{}10^{-15}$\\
     \mr
     $\Gamma^{(d)}_{5}$ &Instrument parameters   & $0.2\times{}10^{-15}$ & $0.1\times{}10^{-15}$ & $0.8\times{}10^{-15}$\\
     \mr
    $ \Gamma^{(d)}_{6}$ &Temperature variations   & $9.3\times{}10^{-15}$ & $17.9\times{}10^{-15}$ & $0.9\times{}10^{-15}$\\
     \mr
     $\Gamma^{(d)}_{7}$ &Drag-Free residuals    & $0.0\times{}10^{-15}$ & $0.0\times{}10^{-15}$ & $0.5\times{}10^{-15}$ \\
     \mr
    $ \Gamma^{(d)}_{8}$ &Magnetic sensitivity   & $0.0\times{}10^{-15}$ & $0.0\times{}10^{-15}$ & $0.4\times{}10^{-15}$ \\
     \mr
    $\Gamma^{(d)}_{9}$ & Non linearity  & $6.0\times{}10^{-15}$ & $3.1\times{}10^{-15}$ & $0.8\times{}10^{-15}$ \\
     \br

\multicolumn{2}{l}{Total quadratic sum (m\,s$^{-2}$)}  &  $11.5\times{}10^{-15}$ & $18.3\times{}10^{-15}$ & \\
     \br
  \multicolumn{5}{l}{\textbf{Total systematic errors for the  E\"otv\"os $\delta$ estimation with $g=7.9$\,m/s$^2$}} \\
  &     Quadratic sum of errors & $1.5\times{}10^{-15}$ & $ 2.3\times{}10^{-15}$  \\
     \br
\end{tabular}
\end{indented}
\end{table}

\section {Conclusion}
Using all scientific and technical sessions, we characterised MICROSCOPE's instrument and its sensitivity to the environment in order to infer the overall systematic error on the estimation of the E\"otv\"os parameter. Thanks to a very performant drag-free satellite and accelerometer, the residual linear and angular accelerations were better than expected. The calibration of the test-mass offcentring and of the instrument differential parameter sensitivity matrix was also performed with better performance than expected prior to mission thanks to the precise determination of the satellite position and orientation in association with the precise GRACE's Earth gravity model. 
The impacts of gravity field variations on board the satellite and magnetic field were estimated with a finite element model associated to on-board thermal environment characterisation.
Dedicated sessions were performed to assess the thermal model in the scientific sessions. The temperature variations in the accelerometer core were evaluated to be lower than 1$\mu$K at  $f_{\rm EP}$ in all science sessions and even one order of magnitude smaller when the satellite is rotating at its maximum rate. 

Temperature sensitivity remains the main contributor followed by the non linearity impact.
Thermal sensitivity and bias are also higher than expected prior to the mission. However, the results of this paper is that the very good satellite environment leads to limit the effect of temperature variations in the systematic error budget. Analysis of specific sessions performed after the publication of first results \cite{touboul17, touboul19} has allowed us to improve the previous MICROSCOPE's EP test systematic errors (in terms of E\"otv\"os ratio) to $1.5\times{}10^{-15}$ for SUEP and to $ 2.3\times{}10^{-15}$ for SUREF. This analysis is essential to provide the MICROSCOPE's final result \cite{metriscqg9} on the  EP test.

\ack

The authors express their gratitude to all the different services involved in the mission partners and in particular CNES, the French space agency in charge of the satellite. This work is based on observations made with the T-SAGE instrument, installed on the CNES-ESA-ONERA-CNRS-OCA-DLR-ZARM MICROSCOPE mission. ONERA authors’ work is financially supported by CNES and ONERA fundings.
Authors from OCA, Observatoire de la C\^ote d’Azur, have been supported by OCA, CNRS, the French National Center for Scientific Research, and CNES. ZARM partners’ work is supported by the DLR, German Space Agency,  with funds of the BMWi (FKZ 50 OY 1305) and by the Deutsche Forschungsgemeinschaft DFG (LA 905/12-1). The authors would like to thank the Physikalisch-Technische Bundesanstalt institute in Braunschweig, Germany, for their contribution to the development of the test-masses with funds of CNES and DLR. 

\appendix
\section {List of acronyms and abbreviations}
DC: Direct Continuous \\
DFACS: Drag-Free and Attitude Control System \\
DFT : Discrete Fourier Transform \\
CGPS: Cold Gas Propulsion System \\
ECM: Electronic Control Module of CGPS \\
FEEU: Front End Electronic Unit of the payload \\
ICUME: Interface Control Unit Mechanical Ensemble of T-SAGE \\
IDEAS: Innovative DEorbiting Aerobrake System \\
MLI: Multi Layer Insulator \\
MT: Micro-Thruster\\
MTB: Magneto Torque Bar\\
OBC: On-Board Computer\\
PCDU: Power Conditioning and Distribution Unit\\
PRM: Pressure Regulation Module\\
Rx/Tx: Receiver/Transmitter\\
STR: Star Tracker\\
SU: Sensor Unit\\
SUEP: Sensor Unit for the Equivalent Principle test\\
SUREF: Sensor Unit for Reference\\
T-SAGE: Twin Space Accelerometer for Gravity Experiment is the name of the payload \\

\section*{References}
\bibliographystyle{iopart-num}
\bibliography{biblimscope}

\end{document}